\documentclass{aa}  
    
    \usepackage{graphicx}
    \usepackage{txfonts}
    \usepackage{booktabs}
    \usepackage{multirow}
    \usepackage{threeparttable}
    \usepackage{tabularx}
    
    \usepackage[colorlinks = true, allcolors=blue]{hyperref}
    \usepackage{orcidlink}
    \usepackage{makecell} 
    \usepackage{amsmath}
    \newcommand\arcdeg{\mbox{$^\circ$}}%

\begin{document}

       \title{UOCS. XVI. Census of post mass-transfer binaries across various evolutionary phases in M67} 

       \titlerunning{Post mass-transfer binaries in M67}
       
       \author{ Harshit Pal\orcidlink{0009-0001-8079-3471}\inst{1,2,5,6},
                Vikrant V. Jadhav\orcidlink{0000-0002-8672-3300}\inst{3,4},
                Arumalla B. S. Reddy\orcidlink{0000-0003-2754-641X}\inst{1},
                Annapurni Subramaniam\orcidlink{0000-0003-4612-620X}\inst{1}}
       
       \institute{
            Indian Institute of Astrophysics, Koramangala II Block, Bangalore-560034, India.
            \and
            Indian Institute of Science Education and Research, Berhampur, Odisha, 760010, India.
            \and
            Astronomical Institute, Faculty of Mathematics and Physics, Charles University, V Holešovičkách 2, CZ-180 00 Praha 8, Czech Republic
            \and
            Helmholtz-Institut für Strahlen- und Kernphysik, Universität Bonn, Nussallee 14-16, D-53115 Bonn, Germany
            \and
            School of Mathematical and Physical Sciences, Macquarie University, Balaclava Road, Sydney, NSW 2109, Australia
            \and
            Astrophysics and Space Technologies Research Centre, Macquarie University, Balaclava Road, Sydney, NSW 2109, Australia\\
            \email{harshit.iiser@gmail.com, vikrant-vinayak.jadhav@matfyz.cuni.cz, balasudhakara.arumalla@iiap.res.in, purni@iiap.res.in}
            }
            
\date{Received March 14, 2025; accepted September 21, 2026}

      \abstract
       {M67 is a well-studied, solar-metallicity, old open cluster that hosts a diverse set of exotic stellar populations formed via mass transfer (MT). While Case A, B, and C MT systems have been individually examined in several clusters, a comprehensive investigation of all MT types across different evolutionary stages within a single cluster remains limited.
       }
       {This study aims to detect the presence of post mass-transfer systems using two different methods and thereby obtain  a holistic view of binary evolution and mass-transfer mechanisms prevalent in M67.
       }
       {
       We constructed spectral energy distributions (SEDs) for 141 UV-detected stars in M67 using far-UV data from AstroSat/UVIT, combined with archival data from Swift/UVOT, GALEX, Gaia, 2MASS, WISE, and other sources. Furthermore, we performed a detailed chemical abundance analysis of cluster members using high-resolution spectroscopy from the GALAH DR3+ survey.
       }
       {
       We identified 11 new MS+WD binaries with previously undetected hotter companions, effectively doubling the known MS+WD binary population in M67. This includes five confirmed ELM WDs with orbital periods <100–200 days and no chemical peculiarities, confirming Case~A or B post-MT systems. We also detected 13 long-period (>1000 days) SB1 systems with enhancements in s-process elements (Ba, La, Y), indicating Case~C post-MT systems enriched by thermally pulsing asymptotic giant branch (AGB) companions which have evolved into undetected WDs.
       }
       {
       This study demonstrates that M67 hosts a diverse population of post-mass transfer binaries formed via Case~A, B, and C mechanisms. The unified view makes M67 a key benchmark for studying binary star evolution and highlights the role of UV observations and high-resolution spectroscopy in uncovering hidden stellar histories.
       }
    

       \maketitle
       \nolinenumbers

\section{Introduction} \label{sec:introduction}

M67, also known as NGC 2682 or the King Cobra cluster, has received significant attention due to its relatively close proximity, with a recent solar-metallicity measurement of  883 $\pm$ 1 pc \citep{dis}, a reddening of E (B-V) = 0.041 $\pm$ 0.004 mag \citep{ev}, and well-identified cluster members by membership studies using both proper motion \citep{sanders,m67ii} and radial velocity measurements \citep{geller}, with various spatial extents and limiting magnitudes. The latest research estimates the age of this solar metallicity open cluster (OC) at $3.95^{+0.16}_{-0.15}$ Gyr \citep{reyes2024isochrone}.
OCs offer a fertile environment for investigating the intricate relationship between binary systems and stellar evolution. Interactions within binary systems have the potential to give rise to stars that defy the conventional model of single-star evolution. M67 is a crucial OC for testing stellar evolution theory due to its proximity, low extinction, well-defined members, large number of solar-type stars, and more.

M67 has been extensively studied through photometry in various multiwavelength bands \citep{m67ccd,m67phot,m67uv,m67ir,mooley}. The cluster contains several unique exotic stellar populations, including blue straggler stars (BSS), sub-subgiants (SSG), yellow straggler stars (YSS), X-ray sources, and triple systems, whose formation processes are not yet fully understood. Being one of the nearest old OCs hosting these above exotica, M67 is indeed a testbed to study non-standard stellar evolution \citep{uocsiii}.  

This cluster hosts a well-studied population of BSSs, identified from many photometric and spectroscopic studies \citep{geller, uocsii, linck2026distribution}. Their diverse nature suggests formation through merger \citep{hills1976stellar} or mass transfer (MT; \citealt{mccrea1964extended}) pathways, specifically via Case~A, B, or C scenarios.
These different MT channels provide crucial insights into BSS formation, where Case~A MT occurs while the donor star is still on the MS, resulting in a binary with a short period \citep{webbink1976evolution}. Case~B MT, which occurs when the donor is on the red giant branch (RGB), results in a system with a helium white dwarf (He WD) companion \citep{mccrea1964extended}. Case~C MT happens after the donor star has evolved into the asymptotic giant branch (AGB; \citealt{ webbink1976evolution}), leading to a longer-period binary and leaving behind a carbon-oxygen (CO) WD \citep{lauterborn1970}. 
However, differentiating between the MT scenarios is non-trivial. Based on the mass of the WD, we can make an educated guess on whether the MT event happened before or after the He core was fully formed. Thus, we use the term `Case~C' for systems with nominal-mass WDs evolved from a fully formed core, and the term `Case~A or B' for low-mass WDs formed via MT, which presumably started before the He core was fully formed.
Understanding these MT pathways is essential for understanding the  diverse populations of BSSs observed in clusters like M67.

In the past, there have been many studies of BSS and other stellar exotica, along with their companions in M67 \citep{m67uv, liu2008spectroscopic,mota, sindhu2019detection, leiner2019blue, uocsii,nine2023wiyn, leiner2025blue}. Here, we present the study of unresolved post MT (Case~A or B) MS-WD binary systems in M67 using similar methods, as well as noteworthy Case~C post-MT binaries, which highlight the diverse pathways of MT at every stage of the star's evolution in the cluster.

In WD-MS binaries, the contrasting spectral energy distributions (SEDs) of the two components provide a method for identification, as the emission profiles are distinct. A hot WD’s emission typically peaks in the UV wavelengths (e.g., FUV-NUV < 3.5 mag; \citealt{finley1989far}), while the MS star exhibits its peak in the optical range. Therefore, an excess UV flux can indicate the presence of a hot companion star in a binary system, as shown in previous studies \citep{vennes1998hot, subramaniam,uocsi, leiner2019blue, uocsviii, nine2023wiyn}. 
Not all post-MT products identified through excess UV flux occupy the classical BSS region of the CMD; those that remain embedded within the cluster MS may instead be classified as blue lurkers \citep{leiner2019blue, Jadhav2026A&A...707A.275J}.
Moreover, Case~C post-MT systems with cooled-down WD companions may escape detection in UV, along with products of mergers or collisions \citep{zorotovic2011post}.

The mass of WD companions depends on the mass of their progenitor stars, with more massive progenitors leading to more massive WDs \citep{cummings}. This relationship is further influenced by factors such as metallicity, rotation, and the binary environment. While single stellar evolution typically yields WDs with a minimum mass of $\sim$ 0.4 $\mathrm{M}_{\odot}$ \citep{kilic}, WDs with masses below 0.3 $\mathrm{M}_{\odot}$ are classified as extremely low-mass (ELM) WDs \citep{elm, li2019formation}. Producing an ELM WD requires stripping the donor’s hydrogen-rich envelope before a substantial helium core develops, which can only occur through early binary interaction \citep{istrate2016}. Such early stripping is consistent with MT beginning during or shortly after the MS phase; however, distinguishing whether Roche-lobe overflow (RLOF) started during core-hydrogen burning (Case~A) or just after hydrogen exhaustion (Case~B) cannot be uniquely inferred from photometry or position in the colour-magnitude diagram (CMD) alone. A complete distinction between these channels requires orbital constraints and comparison with detailed binary-evolution models \citep{wellstein2001,schurmann2024}.

Common-envelope evolution (CEE) provides an additional pathway for forming close (short-period) WD-MS binaries, in which unstable MT leads to envelope ejection and rapid orbital shrinkage \citep{ivanova2013}. Observed WD-MS post-common-envelope binaries (PCEBs; pre-cataclysmic variables) are dominated by orbital periods of hours to a few days, with only a small long-period tail \citep{zorotovic2010,nebot2011orbital}. Such WD-MS systems have also been identified in open clusters through UV and multi-band photometry combined with Gaia astrometry, providing new benchmarks for CEE in cluster environments \citep{grondin2024first,leiner2025blue}.

As M67 has a solar-like chemical composition ([Fe/H] = 0), yet our spectroscopic analysis revealed peculiar chemical behaviours in some stars, including enhancement of s-process elements, indicating these stars have undergone MT \citep{plavec1968mass}.  Such chemical signatures are characteristic of Barium (Ba II) stars, where the primary star has gained enhanced barium and s-process elements through MT from its companion, an AGB star that has since evolved into a WD after Case~C MT scenarios \citep{mcclure1984barium, liu2009abundance}. These stars are known SB1 systems, with longer orbital periods, aiding the verification of Case~C MT. 
\citet{pal2024discovery} reported one such Barium BSS in M67, and provided its detailed analysis. \citet{nine2024wiyn} previously investigated Ba abundances in BSSs in NGC 7789 and M67. In this study, we expand the analysis in M67 by examining the abundances of all s-process elements across various stages of stellar evolution, offering a more comprehensive understanding of the cluster.

The aim of this paper is to take a census of all types of MT (Case~A, B, and C) that can be traced in M67 at various stages of stellar evolution, such as main-sequence stars (MS), blue straggler stars (BSSs), yellow straggler stars (YSSs), subgiants (SGs), and red giants (RGs). We identified Case~A or B MT by detecting the presence of ELM WD companions alongside MS stars using multi-wavelength data. For Case~C MT, we analysed long-period SB1 stars, showing s-process elements enhancement, further supporting the occurrence of all types of MT throughout different phases of stars in this cluster.
The structure of this paper is as follows: Section \ref{sec:data} presents the details of multi-wavelength data, Section \ref{sec:analysis} covers SED fit analysis, Section \ref{sec:characteristics} explores hot companion characteristics, Section \ref{sec:spectral} examines chemical signatures of the stars, and Section \ref{sec:summary} concludes the study.

\section{Data} \label{sec:data}

\subsection{Ultra-Violet Imaging Telescope (UVIT)}

The Ultra-Violet Imaging Telescope (UVIT) is one of the five payloads on the \textit{Astrosat} satellite, the first Indian multi-wavelength space observatory launched on 2015, September 28. The telescope is equipped with three channels,  each containing a set of filters: the FUV, NUV, and VIS channels. The VIS channel is intended to correct the drift of the spacecraft \citep[refer to][for details]{kumar, tandon}.

For this study, we have used the UVIT catalogue of M67 from \citet{uocsiii}. The catalogue includes magnitudes and errors in the three FUV filters: F148W, F154W, and F169M, along with the membership classification (M: member, C: candidate, and F: Field). NUV data was not available due to a technical problem. 
We created a dataset of stars from the \citet{uocsiii} catalogue of M67, including only stars with class M (member), class C (candidate), or undefined class. We excluded stars with class F (field) from our study.

\subsection{ Ultraviolet Optical Telescope (UVOT)}

Launched in 2004, UVOT is one of the three onboard instruments on NASA's \textit{Swift} satellites, focused on UV and optical wavelength observations \citep{siegel1}.
For this study, we utilised archival photometric data from \textit{Swift}/UVOT
in 3 UV filters (UVW1, UVM2, and UVW2), and in 1 optical filter (U; \citealt{siegel3})\footnote{\url{https://archive.stsci.edu/hlsp/uvot-oc}}.

We cross-matched data from UVIT and UVOT catalogues, using a maximum separation of 1.0 arcmin, and created a dataset of 141 stars. WOCS IDs were assigned by integrating with the radial velocities catalogue by \citet{geller}.

\subsection{Archival photometry}

The UVIT and UVOT data are merged with observations from longer wavelengths using VOSA's Virtual Observatory \citep{vosa} to build multi-wavelength SEDs and study the characteristics of all the stars.
We obtained photometric fluxes in the range from UV to IR which includes data from\footnote{All filters used are mentioned at SVO filter profile service: \url{http://svo2.cab.inta-csic.es/theory/fps/}}: GALEX--GR6+7 (FUV and NUV) \citep{galex}, APASS 9 (B,V,g,r,i) \citep{apass}, \textit{Gaia} DR3 (G, Gbp, Grp, Grvs), \textit{Gaia} DR3 J-PAS Synt.Phot. \citep{gaia}, Pan Starrs PS1 DR2 (g, r, i, z, y) \citep{panstars}, J-PLUS DR3 \citep{lopez2024j}, Stroemgren-Crawford uvby photometry catalog (u, v, b, y) \citep{Stroemgren}, SDSS Catalogue, Release 12 (u, g, r, i, z), 2MASS All-Sky Point Source Catalog (J, H, ks) \citep{2mass}, WISE (W1, W2, W3, W4) \citep{wise}. All sources have their extinction (reddening) corrected based on the established extinction-wavelength relations given by \citet{fitzpatrick} and \citet{indebetouw}. We have adopted a distance modulus of 9.6 $\pm$ 0.04 mag \citep{dis}, Solar metallicity, and reddening E (B - V) = 0.041 $\pm$ 0.004 mag \citep{ev}.

We removed the stars with neighbours within 5\arcsec\ to reduce source confusion and flux contamination in archival photometry.
We found WOCS~8009, 6019, 7012, 5019, 3010, 11005, and 1004 to have multiple sources within 5\arcsec. WOCS~7008, 3012, and 4008 are known triple systems \citep{geller} and require a three-component fitting. Therefore, they were also excluded from the SED analysis. 

\subsection{Archival spectroscopy}

We have used the catalogue from \citet{geller2021}, which has RVs of 120 spectroscopic binary populations in M67. The catalogue also provides membership and orbital parameters, from which we have used the relevant orbital data for the members included in our study.

In addition to RVs, we compiled elemental abundances from GALAH+ Survey DR3 \citep{galah} and APOGEE-2 DR16 \citep{apogee}. 
The GALAH+ Survey DR3 provides data from 678,423 optical spectra for 588,571 stars, within 2 kpc, collected using the HERMES spectrograph on the Anglo-Australian Telescope with R=28000. With four discrete optical wavelength channels, it provides detailed stellar parameters and abundance ratios for 30 elements, drawing on data from various observational campaigns and astrometric insights from Gaia DR2 \citep{galah}.

\section{Analysis} \label{sec:analysis}

\begin{table*}\centering
\caption{Parameters obtained by fitting SEDs of stars with UV excesses.
}
\label{tab:SED}
\renewcommand{\arraystretch}{1.2} 
\resizebox{\textwidth}{!}{
\small 
\begin{tabular}{lrrrrrrrrr}\toprule
 & \multicolumn{5}{c}{Cooler Component - A} & \multicolumn{4}{c}{Hotter Component - B} \\\cmidrule{1-10}
S. No. & WOCS  & $T_{\text{eff, A}}$ $(K)$ & $R_{\text{A}}$ ($R_{\odot}$) & $L_{\text{A}}$ ($L_{\odot}$) & $\chi_{r,\text{A}}^2$ & $T_{\text{eff, B}}$  $(K)$ & $R_{\text{B}}$ ($R_{\odot}$) & $L_{\text{B}}$ ($L_{\odot}$) & $\chi_{r,\text{B}}^2$ \\\midrule

\multicolumn{10}{c}{\textbf{Kurucz + Koester}} \\

1  & 5013  & $6250^{+250}_{-250}$ & $1.66^{+0.03}_{-0.03}$ & $3.78^{+0.7}_{-0.8}$ & $9.87$ & $11250^{+8272}_{-2279}$ & $0.032^{+0.001}_{-0.0008}$ & $0.018^{+0.143}_{-0.025}$ & $14.47$ \\
2  & 12010 & $6250^{+250}_{-250}$ & $1.43^{+0.03}_{-0.03}$ & $2.8^{+0.6}_{-0.5}$ & $8.52$ & $11250^{+6290}_{-2470}$ & $0.0421^{+0.0007}_{-0.002}$ & $0.023^{+0.389}_{-0.025}$ & $8.64$ \\
3  & 5018  & $6250^{+250}_{-250}$ & $1.39^{+0.02}_{-0.02}$ & $2.6^{+0.6}_{-0.5}$ & $7.16$ & $11500^{+3760}_{-1829}$ & $0.0457^{+0.0008}_{-0.0008}$ & $0.027^{+0.187}_{-0.047}$ & $8.38$ \\
4  & 6022  & $6250^{+250}_{-250}$ & $1.69^{+0.03}_{-0.03}$ & $3.9^{+0.7}_{-0.8}$ & $19.09$ & $13250^{+8786}_{-2711}$ & $0.009^{+0.001}_{-0.001}$ & $0.004^{+0.097}_{-0.009}$ & $19.27$ \\
5  & 6005  & $5500^{+250}_{-250}$ & $2.90^{+0.05}_{-0.05}$ & $6.94^{+1.4}_{-1.6}$ & $23.98$ & $10500^{+2074}_{-1006}$ & $0.086^{+0.006}_{-0.0015}$ & $0.070^{+0.071}_{-0.039}$ & $33.31$ \\
6  & 2002  & $5250^{+250}_{-250}$ & $5.4^{+0.1}_{-0.1}$ & $20^{+4.0}_{-5.0}$ & $13.28$ & $14500^{+1639}_{-1573}$ & $0.054^{+0.001}_{-0.001}$ & $0.13^{+0.037}_{-0.023}$ & $20.08$ \\
7  & 7002  & $5250^{+250}_{-250}$ & $1.48^{+0.03}_{-0.03}$ & $1.5^{+0.3}_{-0.4}$ & $20.38$ & $11000^{+2432}_{-2077}$ & $0.0313^{+0.0005}_{-0.0005}$ & $0.016^{+0.015}_{-0.006}$ & $26.82$ \\
8  & 9011  & $6000^{+250}_{-250}$ & $1.65^{+0.03}_{-0.03}$ & $3.2^{+0.6}_{-0.7}$ & $41.82$ & $11500^{+2713}_{-1775}$ & $0.091^{+0.002}_{-0.002}$ & $0.08^{+0.177}_{-0.051}$ & $85.23$ \\
9  & 2012  & $6000^{+250}_{-250}$ & $2.22^{+0.04}_{-0.04}$ & $5.8^{+1.1}_{-1.3}$ & $17.92$ & $11000^{+4728}_{-1490}$ & $0.0495^{+0.0009}_{-0.0009}$ & $0.0354^{+0.108}_{-0.042}$ & $34.91$ \\
10 & 7022  & $6000^{+250}_{-250}$ & $1.66^{+0.03}_{-0.03}$ & $3.78^{+0.1}_{-0.1}$ & $4.84$ & $10000^{+9056}_{-3172}$ & $0.0367^{+0.001}_{-0.001}$ & $0.0122^{+0.074}_{-0.003}$ & $5.09$ \\

\midrule
\multicolumn{10}{c}{\textbf{KuruczUVBlue + Koester}} \\

11 & 4015 & $6000^{+500}_{-500}$ & $2.11^{+0.04}_{-0.04}$ & $5.19^{+0.14}_{-0.14}$ & $23.4$ & $9750^{+2409}_{-1102}$ & $0.124^{+0.0022}_{-0.0022}$ & $0.126^{+0.133}_{-0.090}$ & $44.45$ \\
12 & 4014 & $6000^{+500}_{-500}$ & $2.23^{+0.04}_{-0.04}$ & $5.82^{+0.15}_{-0.15}$ & $13.97$ & $9250^{+4297}_{-1342}$ & $0.105^{+0.003}_{-0.008}$ & $0.071^{+0.656}_{-0.155}$ & $16.93$ \\
13 & 3001 & $6500^{+500}_{-500}$ & $1.40^{+0.02}_{-0.02}$ & $3.14^{+0.09}_{-0.08}$ & $3.62$ & $11250^{+12310}_{-2722}$ & $0.0562^{+0.0038}_{-0.0015}$ & $0.0417^{+0.540}_{-0.076}$ & $5.20$ \\
14 & 4002 & $6000^{+500}_{-500}$ & $1.70^{+0.03}_{-0.03}$ & $3.36^{+0.09}_{-0.15}$ & $26.46$ & $10000^{+3919}_{-1735}$ & $0.0646^{+0.0011}_{-0.0017}$ & $0.0457^{+0.186}_{-0.046}$ & $34.67$ \\
15 & 5002 & $6000^{+500}_{-500}$ & $1.51^{+0.03}_{-0.03}$ & $2.67^{+0.07}_{-0.16}$ & $14.01$ & $10500^{+7223}_{-1359}$ & $0.0692^{+0.0018}_{-0.0012}$ & $0.0431^{+0.176}_{-0.071}$ & $28.53$ \\

\bottomrule
\end{tabular}}
\tablefoot{The parameters above the line in the table are obtained from Kurucz + Koester models, while those below the line are from KuruczUVBlue + Koester binary fits. The best-fit parameters among all models are presented in this table. The full version of this table at the CDS covers all fitted stars and also includes the UVIT and UVOT photometry, all SED-fitting parameters for the different models, and other data from this study.}
\end{table*}

\subsection{Spectral energy distributions}
\label{subsec:SED}

Advancements in observational facilities spanning a broad range of electromagnetic wavelengths have enabled us to create spectral energy distributions (SEDs) spanning from UV to IR wavelengths, which serve as a valuable tool for determining the atmospheric characteristics of stellar objects, including their effective temperature ($T_{eff}$), radius (R), and luminosity (L). Moreover, an excess in UV fluxes can suggest the presence of an optically faint, hotter companion of that star. 
To fit the SEDs, we carried out the following steps:

\begin{enumerate}
    \item We used virtual observatory SED analyzer (VOSA; \citealt{vosa}) to construct SEDs of all the sources. We used the updated Kurucz stellar models \citep{castelli} with the parameters of [Fe/H] = 0, log g = 3--5 dex, and $T_{eff}$ = 3500--50000 K for fitting all data above 3000 \AA. From each fit, we obtained the corresponding parameters like $T_{eff}$, radius, and luminosity for the star.
    
    \item We identified 22 Stars with 3 or more data points in their single SED fit with fractional residual, $(flux_{obs}-flux_{fit})/flux_{obs} > 0.5$ in the UV region and performed a binary double-component fit on all of them. This threshold is the UV-excess criterion needed to justify a two-component binary SED fit. The double-component fit of these 22 stars was done using a Python-based \textsc{Binary SED Fitting}\footnote{\url{https://github.com/jikrant3/Binary_SED_Fitting}} code \citep{Jadhav2021JApA...42...89J}.
    
    \item The cooler components were fitted with either a Kurucz model or a Kurucz\_UBVLUE hybrid model following the method in \citet{jadhav2023uocs}. The hybrid model consists of Kurucz spectra for 4700 \AA\ and above while using UVBLUE spectra \citep{rodriguez2005uvblue} for the 850--4700 \AA\ range. The UVBLUE models were convolved with UVIT filters  \citep{tandon2020additional} and UVOT filters \footnote{\url{https://heasarc.gsfc.nasa.gov/docs/heasarc/caldb/swift/}} to obtain synthetic photometry across a temperature range of 3000 to 50000 $K$ and log g range of 0 to 5, using filter transmission profiles from the SVO Filter Profile Service \citep{2012Rodrigo, 2020Rodrigo}. 
    
    \item The candidate hotter components were fitted with Koester spectra \citep{koester} with parameter range of log g = 6.5--9.5 and $T_{eff}$ = 8000--80000 K.
\end{enumerate}

Out of 22 stars with UV excess, we were able to make satisfactory binary fits for 10 stars with Kurucz and Koester models and for 5 stars with the hybrid Kurucz\_UVBlue and Koester models (see Figure~\ref{fig:SED}). However, we couldn't fit seven stars: WOCS~5004, 1005, 4003, 7007, 12006, 8020, and 8011 using any of the models (see Section~\ref{sec:HRD}). 

We use the approach of \citet{2025JApA...46...20J} to estimate the errors of binary SEDs using \textsc{sed-analysis-tools}\footnote{\url{https://github.com/jikrant3/sed-analysis-tools}}. A synthetic blackbody binary was created based on the results of SED fitting. Gaussian noise was added to the synthetic binary scaled according to the observational flux errors. 100 random realisations of these synthetic binaries were then fitted with an analytical 2-component blackbody SED, and the stellar parameters were recovered. The 16th and 84th percentile values of the recovered parameters (and the model stepsize) were used to estimate the parameter errors\footnote{The errors estimated using this method can be significantly larger than formal SED fitting errors. For example, the fitting errors in the $T_\text{eff,B}$ using the approach used in VOSA and previous UOCS papers have a maximum value of 500 K.}.

The best parameters obtained by fitting the SEDs using the Kurucz + Koester models and the KuruczUVBlue + Koester models are provided in Table~\ref{tab:SED}.
All extended stellar parameters of the cooler and hotter components from the single and binary SED analysis of all stars with all models are given as supplementary material in CDS.

\section{Characteristics of the hot companions} \label{sec:characteristics}

\begin{table*}
\caption{Photometric properties of the candidate hotter companions.}
\label{table:dynamic}
\centering
\renewcommand{\arraystretch}{1.35}
\setlength{\tabcolsep}{6pt}
\begin{tabular}{lcccccccc}
\hline\hline
WOCS & Type & $M_1$ & $M_{2,\min}$ & WD type & Photometric M$_{WD}$ $M_\odot$ & log(age/yr) & SED vs dynamical \\
&&($M_\odot$)&($M_\odot$)&(from $M_{2,\min}$)&(Mean$\pm$errors)&(Mean$\pm$errors)& consistency \\
\hline
12010$^*$ & SB1 & 1.131 & 0.340 & He WD likely        & $0.176_{-0.026}^{+0.824}$ & $9.256_{-0.837}^{+0.773}$ & ELM WD ruled out \\
5018$^\dagger$   & SB1 & 1.117 & 0.192 & ELM-He WD possible & $0.161_{-0.006}^{+0.042}$ & $9.253_{-0.834}^{+0.426}$ & Yes \\
2002$^*$  & SB1 & 1.326 & 0.229 & ELM-He WD possible & $0.182_{-0.002}^{+0.008}$ & $9.037_{-0.233}^{+0.002}$ & Not consistent \\
3001$^*$  & SB1 & 1.149 & 0.330 & He WD likely        & $0.165_{-0.010}^{+0.835}$ & $9.138_{-0.719}^{+0.665}$ & ELM WD ruled out \\
7022$^*$  & SB1 & 1.194 & 0.234 & ELM-He WD possible & $0.500_{-0.350}^{+0.800}$ & $8.346_{-2.782}^{+1.193}$ & Yes \\
9011             & SB1 & 1.210 & 0.703 & CO WD likely        & $0.182_{-0.027}^{+0.020}$ & $8.977_{-0.558}^{+0.488}$ & ELM WD ruled out \\
6005             & SB2 & 0.755 & 0.661 & CO WD likely        & $0.161_{-0.011}^{+0.019}$ & $9.208_{-0.724}^{+0.172}$ & ELM WD ruled out \\
7002$^*$  & SB2 & 0.013 & 0.011 & unconstrained      & $0.200_{-0.030}^{+0.300}$ & $6.840_{-0.611}^{+2.439}$ & Yes \\
5013$^*$  & --- & 1.189 & ---   & ---                & $0.176_{-0.026}^{+0.824}$ & $9.249_{-0.830}^{+0.617}$ & --- \\
6022$^*$  & --- & 1.189 & ---   & ---                & $0.200_{-0.050}^{+1.000}$ & $7.042_{-1.478}^{+2.737}$ & --- \\
2012$^\dagger$   & --- & 1.277 & ---   & ---                & $0.164_{-0.009}^{+0.039}$ & $9.223_{-0.804}^{+0.262}$ & --- \\
4015$^\dagger$   & --- & 1.267 & ---   & ---                & $0.182_{-0.027}^{+0.021}$ & $8.870_{-0.451}^{+0.487}$ & --- \\
5002$^*$  & --- & 1.117 & ---   & ---                & $0.182_{-0.032}^{+0.618}$ & $8.959_{-0.540}^{+0.979}$ & --- \\
4002$^\dagger$   & --- & 1.189 & ---   & ---                & $0.161_{-0.006}^{+0.042}$ & $9.233_{-0.814}^{+0.310}$ & --- \\
4014$^\dagger$   & --- & 1.277 & ---   & ---                & $0.182_{-0.027}^{+0.021}$ & $8.950_{-0.530}^{+0.648}$ & --- \\
\hline
\end{tabular}
\tablefoot{
Column 1: WOCS ID of the star. 
Column 2: Spectroscopic binarity \citep{geller2021}.
Column 3: Primary mass derived by fitting the Gaia DR3 photometry (G, BP–RP) to the PARSEC isochrone (Section~\ref{sec:HRD}). For SB2 systems (WOCS~6005 and 7002), photometric mass estimates are unreliable due to their deviation from single-star isochrones; therefore, we adopt the SB2 orbital solutions from \citet{geller2021}, which provide direct dynamical mass constraints ($m_1\sin^3 i$ and $m_2\sin^3 i$).
Column 4: Minimum dynamical mass of the secondary from the spectroscopic orbit-based mass function from \citet{geller2021}.
Column 5: WD companion type inferred from $M_{2,\min}$.
Column 6: Photometric WD masses calculated for all UV-excess binaries with successful SED fitting. We used WD cooling tracks from \citet{WD} for masses in the range $0.2$–$1.3,M_\odot$ and from \citet{elm} for $<0.2,M_\odot$, as shown in Figure~\ref{fig:HRD} for the mass estimation.
Column 7: Photometric age of the WDs using comparison to the cooling models.
Column 8: Consistency between the WD masses from SED fitting and dynamical minimum masses.\\
\tablefoottext{$\dagger$}{Confirmed MS+ELM WD binaries with no X-ray emission detected.}
\tablefoottext{$*$}{Binaries with successful SED binary fits but unconfirmed WD types; their WD mass uncertainties span multiple WD-type regimes (see appendix~\ref{app:UV_excess}).}
}
\end{table*}

\subsection{Hertzsprung–Russell diagram and WD mass and age estimation}\label{sec:HRD}

\begin{figure}
	\includegraphics[width=\columnwidth]{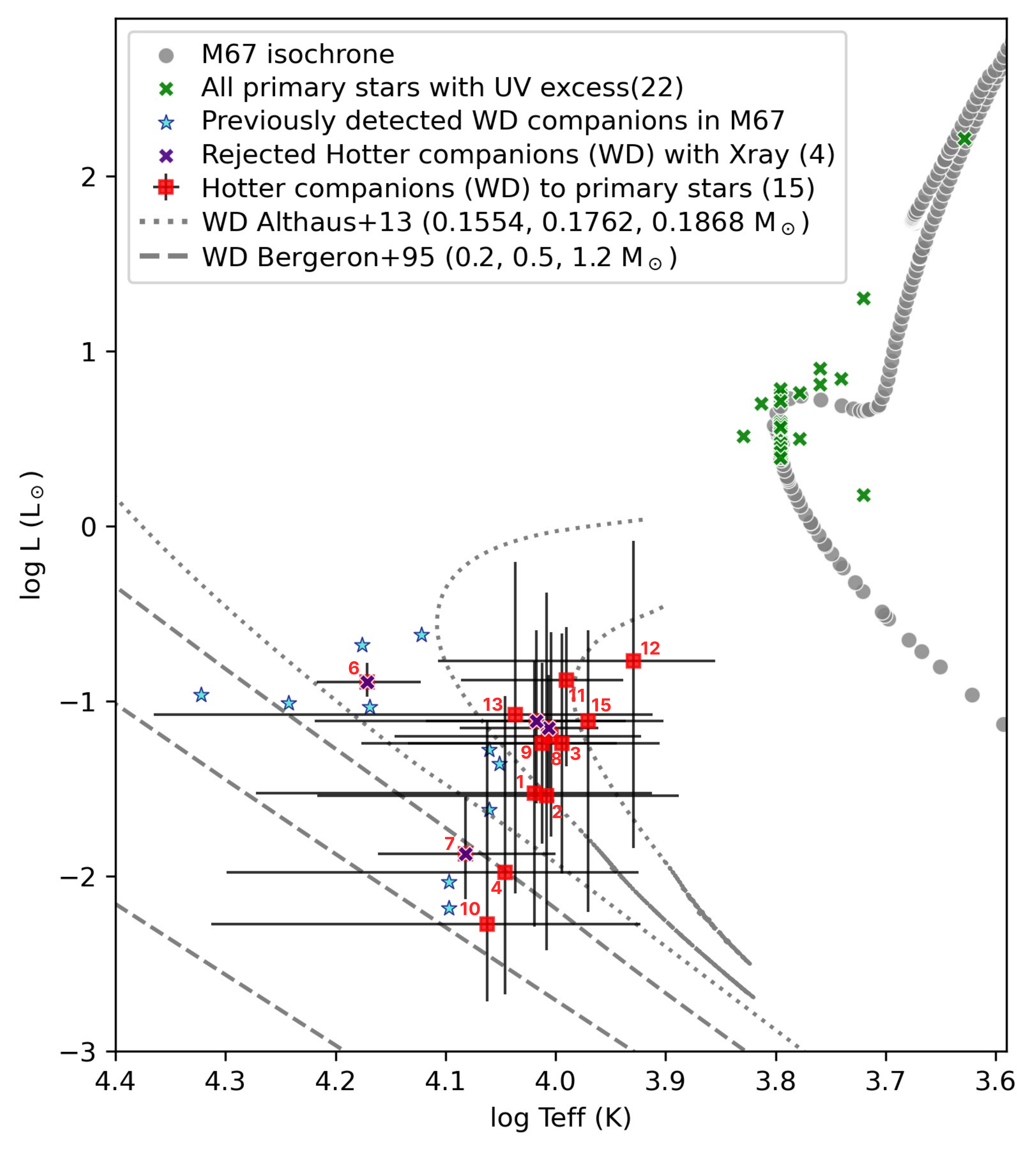}
    \caption{The HRD positions of M67 members and their companions. 
    Green bold crosses denote all 22 members showing significant UV excess and binary fits, while hot companion candidates for these stars, which were successfully fitted, are marked with red-filled squares with black error bars and are labelled by their respective S.No. from Table~\ref{tab:SED}. For all primary star parameters, the fitting errors (max of binary SED fitting error and SED analysis tools error) have step sizes of $\pm$ 250 K (Kurucz) and $\pm$ 500 K (KuruczUVBlue). These fitting errors are not shown in the HRD. Dark purple crosses indicate sources detected by X-ray. Previously identified WD companions in M67 from the literature are marked with cyan stars. The M67 PARSEC isochrone is shown as grey scattered points, while WD cooling curves from 0.2 to 1.3 $\mathrm{M}_{\odot}$ using \citet{WD} and ELM WD sequences with 0.15540 to 0.1868 $\mathrm{M}_{\odot}$ from \citet{elm} are shown as grey dashed and dotted curves, respectively. 
    } 
    \label{fig:HRD}
\end{figure}

We plotted the Hertzsprung–Russell diagram (HRD; Figure~\ref{fig:HRD}) to compare the properties of the hotter components to better understand their nature.
We have overlaid the PARSEC isochrone\footnote{\url{http://stev.oapd.inaf.it/cgi-bin/cmd}} of log (Age) = 9.6, MH = 0.0, and Z = 0.01476  in the HRD for reference.
In this HRD, we have plotted 22 primary stars with UV excess and 15 hotter companions identified using the SED method, along with cooling curves for WDs.
The WD cooling curves provide information on the luminosity, temperature, and log\,g of WDs of various masses and ages. We used \citet{WD} WD models\footnote{\url{http://www.astro.umontreal.ca/~bergeron/CoolingModels}} for DA-type (hydrogen atmosphere) WDs ($0.2-1.3 \mathrm{M}_{\odot}$) and \citet{elm} WD models for He-core ELM WDs ($ \sim 0.15-0.19 \mathrm{M}_{\odot}$).

Seven of these objects exhibited significant UV excesses (Section~\ref{subsec:SED}.2), however, their binary SED fits yielded unphysical or poorly constrained WD parameters (e.g. mass $>$ 1.3 $\mathrm{M}_{\odot}$). Similar UV excesses are known in M67 and other old open clusters to arise from chromospheric activity or other non-degenerate UV sources rather than hot WD companions, and need further study to confirm their nature.
Therefore, these seven WD candidates with HRD masses greater than 1.3 $\mathrm{M}_{\odot}$ were excluded from further analysis. 
The differences between previous WD parameters and the current work are likely due to the usage of the KuruczUVBlue models (which can reduce the UV excess) and updated SED fitting process.
The failed fits also suggest that some apparently successful fits may be spurious; however, quantifying this bias requires dedicated follow-up observations and is beyond the scope of this work. Independent evidence of binarity and RV variability is essential to confirm the nature of all SED based binary components.

We estimated the photometric masses and ages of the WDs by comparing the HRD positions with respect to WD cooling curves. 
Using \textsc{sed-analysis-tools}, we found large uncertainties in the $T_{eff}$ and $L$ parameters of the hotter companions in binary SEDs. As a result, the WD mass and age estimates from the HRD show a wide range of possible solutions.
ELM WDs are WDs below 0.3 $\mathrm{M}_{\odot}$ \citep{brownelm}, WDs range from 0.3 to 0.45 $\mathrm{M}_{\odot}$ are commonly referred to LM WDs \citep{pelisoli2019gaia}, normal-mass WDs range from 0.45 to 0.75  $\mathrm{M}_{\odot}$, and high-mass WDs are defined as >0.75  $\mathrm{M}_{\odot}$ \citep{tremblay2016field}. 

We have verified our photometric WD mass estimates against dynamical lower limits derived from available spectroscopic orbits. The uncertainty corresponds to the $+84\%$ confidence interval on the photometric WD mass from the above analysis, and ELM consistency reflects whether this interval lies entirely below the ELM mass threshold when compared with the dynamical lower limits from available spectroscopic orbits, as summarized in Table~\ref{table:dynamic}.

The inclusion of dynamical constraints substantially refines our sample. 
In particular, 12010, 9011, and 3001 have $M_{2,\min}$ well above the ELM regime, disfavoring ELM companions. 
The SB2 systems (6005, 7002) are likewise removed from the confirmed set. WOCS~2002 is also excluded because its SED-derived WD mass is lower than the dynamical minimum companion mass, and it exhibits X-ray emission (see Section~\ref{subsec:xray} and Table~\ref{table:wd_candidates}). It is therefore marked as inconsistent in Table~\ref{table:dynamic}.
After this consistency check, only nine systems (WOCS~5018, 7022, 5013, 6022, 2012, 4015, 5002, 4002, 4014) remain eligible for MS+ELM WD binaries.

Among these nine systems, only five systems (5018, 2012, 4015, 4002, 4014) have WD masses whose mass range lies entirely within the ELM-type WD and have no inconsistency with dynamical parameters. The other 10 systems, which have large uncertainties in their WD mass estimates, are therefore removed from the list of confirmed post Case~A or B MT binaries. The WD mass and age estimates from HRD analysis are now presented in two tables: Table~\ref{table:dynamic} for the five confirmed systems and Table~\ref{table:wd_candidates} for the remaining ten systems.

\subsection{X-ray emission}\label{subsec:xray}
X-ray emission on a star can suggest surface activity, such as hot spots or coronal activity, which can contribute to the UV flux and make it difficult to determine if a hotter companion exists. If X-ray emission is detected, it may also suggest the presence of additional sources of high-energy radiation amounting to the UV flux besides the companion star, which could be related to the star's own magnetic activity or accretion from a circumstellar disk \citep{mitra}. However, even if there is a hotter companion, the $T_{eff}$ and radius obtained via SED fitting may not be accurate, making it difficult to comment on the presence of a WD companion \citep{uocsiii}. 
X-ray observation data (in Table~\ref{table:wd_candidates}) were taken from the archival XMM-Newton data catalogue by \citet{mooley} and from a 47-ks Chandra-ACIS observation by \citet{Van}.
Thus, X-ray detected systems (such as those listed in Table~\ref{table:wd_candidates}), despite having good SED binary fits and acceptable errors, cannot be fully confirmed to host hotter WD companions.

\subsection{Relation between WD temperature and rotation}

We compared the temperature of the ELM WDs (hotter companions) against the v$\sin{i}$ of binary systems derived from our own spectral analysis of GALAH data (see Section~\ref{sec:spectral} and Table~\ref{tab:params} for details). The v$\sin{i}$ value of a star, which is the projected rotational velocity measured from its spectral lines, provides important insights into its age, mass, magnetic field strength, and MT in binary systems \citep{diaz2011accurate}. Recent MT can impart spin in the accreating star through angular-momentum accretion \citep{packet1981, deMink2013ApJ...764..166D, matrozis2017}, however the rotation signal is diminished if the system is pole-on or the MT occurred a while ago. 
Although rapid rotation has been used as an observational indicator of recent MT, no correlation was found between the two in this study.

\begin{figure*}
\sidecaption
\includegraphics[width=12.9cm]{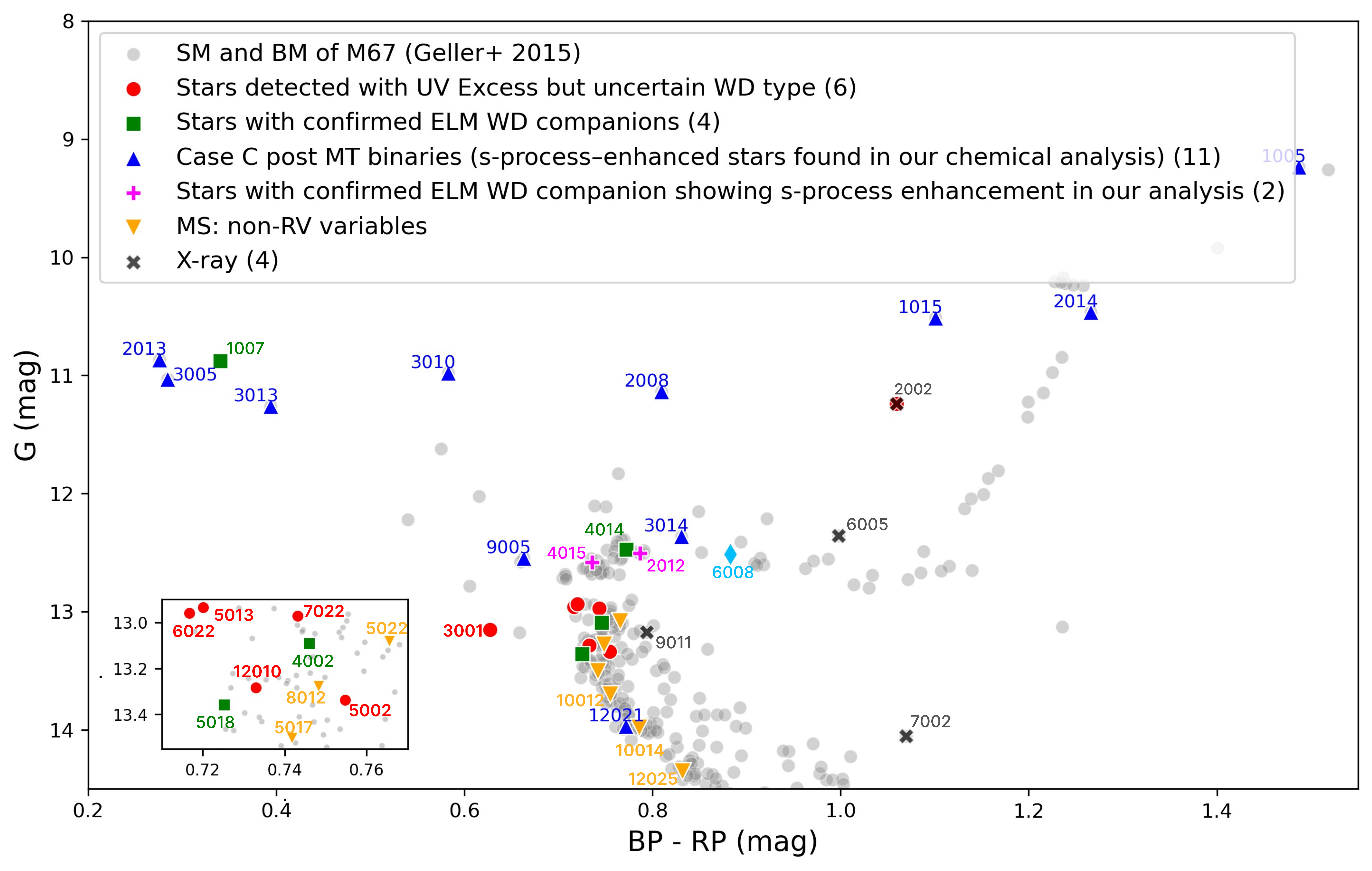}
\caption{Optical CMD of single and binary members of M67 from \citet{geller}. Stars with WD companions (from SEDs; Sect.~\ref{sec:characteristics}) are marked with red circles for uncertain WD types (Table~\ref{table:wd_candidates}) and green squares for confirmed ELM WD companions (Table~\ref{table:dynamic}), whereas X-ray detected stars are marked by black crosses. Blue triangles indicate stars with s-process enhancement identified from our independent chemical analysis of each star (see Sect.~\ref{sec:spectral}, Table~\ref{tab:abundance}). WOCS~4015 and 2012 (shown as magenta plus signs) both host a confirmed ELM WD companion and also show s-process enhancement in our analysis (Sect.~\ref{sec:spectral}).}
\label{fig:CMD}
\end{figure*}

\subsection{Optical colour--magnitude diagram (CMD)}

The CMDs are also crucial for classifying stars in different evolutionary stages. Figure~\ref{fig:CMD} shows the positions of cooler components alongside the M67 members from \citet{geller}. Binaries listed in Table~\ref{table:dynamic} with ELM WD companions are shown, where many are found in the MS. As these are post-MT systems, they are blue lurker candidates. 
For completeness in the CMD, we have also included WOCS~1007, a previously identified BSS with an ELM WD companion based on dynamical studies and our previous UOCS studies (further details in Section~\ref{sec:discussion}). WOCS~6008 is shown as a light blue diamond in Fig~\ref{fig:CMD}; it is a short-period SB1 system with a strong FUV excess previously detected using GALEX photometry \citep{subramaniam}. We retain it in this census as a UV-excess post-MT binary system, and present a detailed spectral analysis in Section~\ref{sec:spectral}.
The remaining six UV-excess MS + WD systems listed in Appendix~A (without X-ray emission detection) are also shown in the CMD, but their WD types still remain unconstrained. We note that those with ELM companions are found in the MS, BSS region and the sub-giant region.

However, the ELM identification and the MS location of the primaries in the CMD alone do not uniquely distinguish between Case~A and Case~B channels; a solid classification requires orbital parameters and modelling, as is standard in binary-evolution studies of ELM and low-mass He-core WDs \citep[e.g.][]{sun2018,mathieu2025blue}.

\section{Abundances in post mass transfer systems}\label{sec:spectral}

In order to gain a deeper understanding of the effects that binary interactions and subsequent mass transfer can have on the evolution of stars, we examined the elemental abundance information of M67 stars reported by the GALAH survey \citep{galah}. Of particular interest is the identification of stars showing enhancements in $s$-process elements relative to the mean cluster abundance, if the MT (Case~C) from the thermally pulsing AGB companion is responsible for the formation of some of the BSSs in M67.

\begin{figure*}
    \centering
    \includegraphics[
        width=\textwidth
    ]{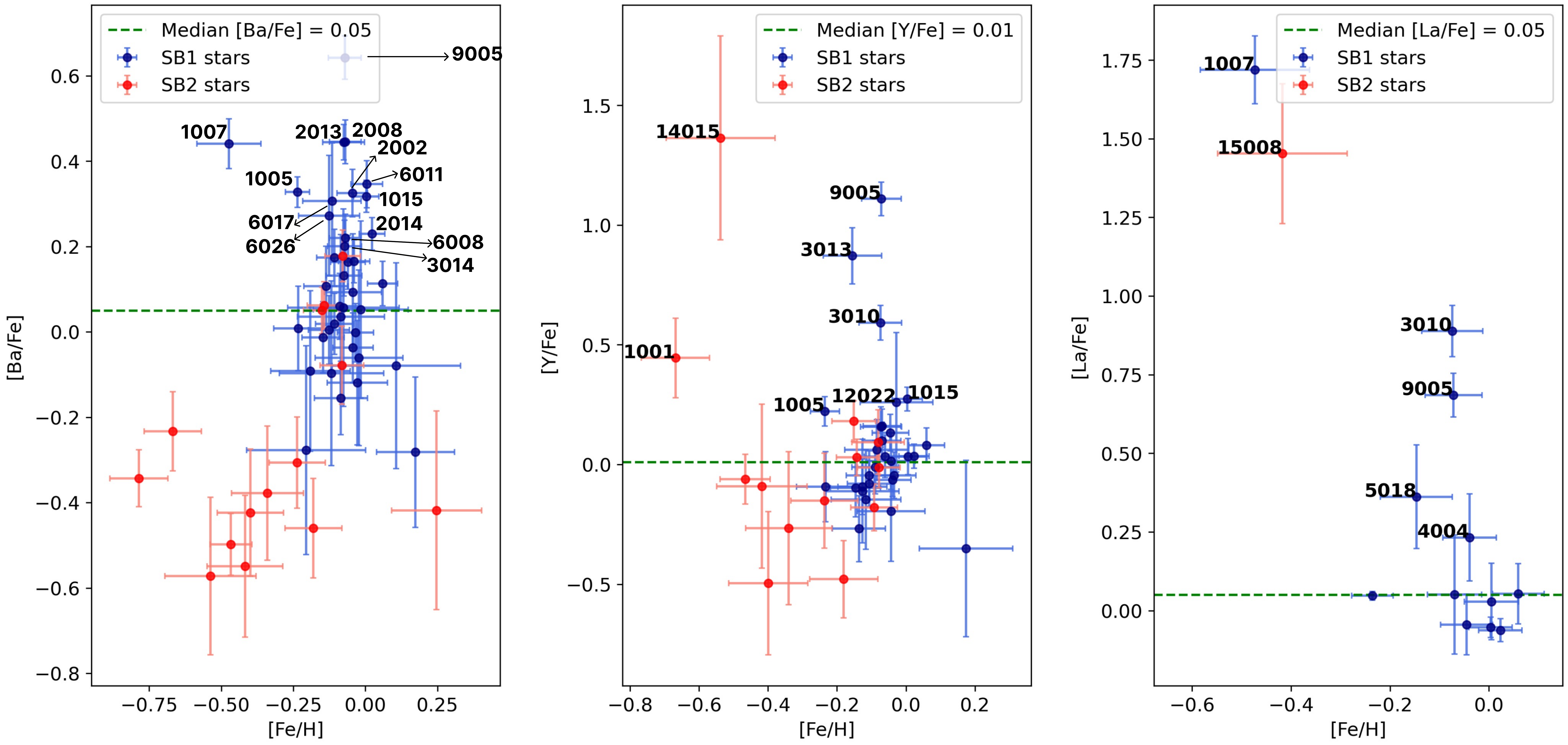}
    \caption{[X/Fe] versus [Fe/H] for all M67 spectroscopic binary members using GALAH DR3 data \citep{galah}, where X refers to the s-process elements Ba, Y, and La drawn from the automated pipeline catalogue.}
    \label{fig:abundance}
\end{figure*}

Fig \ref{fig:abundance} shows the [X/Fe] vs [Fe/H] for the abundances of X$=$ Ba, Y, and La drawn from the automated pipeline measured values from the GALAH DR3 survey \citep{galah}. Although the mean metallicity of the cluster M67 is close to solar as measured for dwarfs \citep{nine2024wiyn} and red giants (see Table 13 in \citealt{reddy2013comprehensive}), [Fe/H] values for many stars in the figure ([Ba/Fe] vs. [Fe/H]) deviate from the cluster mean [Fe/H] by more than 2-$\sigma$ error.
Inspection of the spectra reveals that the pipeline-measured abundances are negatively affected by a multitude of issues related to spectral extraction, low signal-to-noise ratio (SNR), unresolved binarity (SB2), and rapid rotation of the star. The typical SNR of the observed spectra falls in the range of 2 to 200, with a vast majority of stars analysed for abundances having SNR $< 40$. Therefore, caution is advised when interpreting these results at face value \citep{gaia}. 

Thus, we performed an independent chemical composition analysis for the SB1 stars marked in Fig~\ref{fig:abundance}. 
Along with these, we also selected a control sample of seven MS stars to establish the baseline of the cluster's mean abundance. These MS stars are non–RV variables, and non-rapid rotators ensuring reliable abundance determinations.  The optical spectra of these stars were retrieved from the GALAH DR3 Spectral Library\footnote{\url{https://www.galah-survey.org/dr3/the_spectra/}}. The stellar parameters and chemical abundances for the key lines of interest (Y, Ba, and La) were measured in a similar way as outlined in \citet{pal2024discovery}. The list of stars analysed, as well as the derived stellar parameters along with $v\sin i$ values, are tabulated in Table \ref{tab:params}.

The elemental abundances listed in Table \ref{tab:heavy_elements_galah_vs_thisstudy} for all stars are derived from the spectrum synthesis of singly ionized lines: yttrium at $\lambda\,4884$ \AA, barium at $\lambda\lambda\,5853$ and $6496$ \AA\, and lanthanum at $\lambda\,5806$ \AA. For WOCS~12021, however, the barium abundance is determined solely from the barium line at $\lambda\,6496$ \AA, as the line at $\lambda\,5853$ \AA\ is unusable. The spectrum synthesis of the $\lambda\,6496$ \AA\ line yields [Ba/Fe] $= 0.35\pm0.08$ dex for this star, indicating a significant barium enhancement.   Table~\ref{tab:heavy_elements_galah_vs_thisstudy} lists the abundances and typical 1-$\sigma$ errors derived in this study alongside those obtained from the GALAH pipeline (in Fig~\ref{fig:abundance}). The mean abundances of the control sample are [Fe/H] = $-0.05\pm0.06$ dex, [Y/Fe] = $+0.06\pm0.08$ dex, [Ba/Fe] = $+0.05\pm0.10$ dex, and [La/Fe] = $0.00\pm0.08$ dex. A comparison of the MS-star abundances with those listed in Table~\ref{tab:abundance} identifies 13 stars exhibiting clear s-process elemental enrichment, with abundance enhancements exceeding the 3$\sigma$ of the MS baseline abundances, indicative of a past episode of Case~C MT in a binary system with an AGB companion that has since evolved into an unseen WD. The locations of these stars in the CMD are shown in Figure~\ref{fig:CMD}. It can be seen that these s-process–enhanced stars are not confined to any particular region of the CMD but span the MS, BSS, and evolved populations, demonstrating that the present-day products of past MT are found across a wide range of evolutionary stages in M67. Thus, although the evolutionary stage at which MT occurred cannot be inferred from the present-day CMD position alone, its chemical imprint persists across the evolutionary landscape of M67.

While the long orbital periods of most s-process–enriched stars are consistent with Case~C MT from an AGB companion (e.g. \citealt{hurley2002,chen2008mass}), a few s-process–enriched stars (WOCS~2012, 4015, and 3005) show no detectable RV variations. Current theoretical models for Ba enrichment outside the AGB phase generally invoke MT from a thermally pulsing AGB companion, and observations of barium giants and dwarfs in the field support that over 85\% of such stars are members of binary systems \citep{escorza2019barium}. It is also possible that these stars are members of binary systems with similarly long orbital periods that remain undetected by current RV measurements or that the binary was disrupted post-interaction. This apparent discrepancy may be reconciled if the MT occurred via stellar winds or wind Roche-lobe overflow (WRLOF) mechanisms that allow chemically enriched material from an AGB donor to be accreted onto the primary through stellar winds.

The detection of ELM WD companions to MS stars (MS+ELM WD system) indicates a mass transfer history, but not all these stars exhibit detectable elemental enhancement (e.g. WOCS~2012 and 4015 exhibit a clear $s$-process enhancement). Similar behaviour is observed among the SGB and YSS candidates. This diversity may reflect the accretion of largely unprocessed material (Case-A MT) and/or a transient nature of surface overabundances following Case-B/C MT. In contrast, stars identified as long orbital period spectroscopic binaries by \citet{geller2021} exhibit chemical enhancements of varying magnitudes, yet no evidence for hotter companions is found from the SED fitting of excess UV flux. This suggests that these systems likely host white dwarf companions that have cooled and faded sufficiently to evade detection in the UV.

The identification of ELM WD companions to host stars lacking $s$-process enrichment, together with confirmed $s$-process–enriched stars both with/without hotter companions, and with all host stars distributed across the CMD rather than confined to a single evolutionary phase, provides strong evidence that M67 hosts binary systems that have experienced all canonical mass-transfer channels: Case~A, Case~B, and Case~C. A definitive assessment of the dominant mass-transfer mechanism responsible for binary system formation in M67 will require a detailed analysis of a statistically significant sample with well-determined CNO and $s$-process elemental abundances. We defer such an investigation to a future study.

\begin{table}
\centering
\caption{Stellar parameters for a sample of confirmed binary members and non-RV variables (control sample) of M67 derived in this study. }
\label{tab:params}
\begin{tabular}{lcccc}
\toprule
WOCS & $T_{\mathrm{eff}}$ (K) & $\log g$ & $v_{\mathrm{mic}}$ & $v\sin i$ \\
\midrule

\multicolumn{5}{l}{\textbf{MS}}\\
2012  & 5875 & 3.8 & 1.2 & 6  \\
3001  & 6400 & 4.2 & 1.5 & 23 \\
4002  & 6000 & 4.0 & 1.1 & 9  \\
4014  & 6006 & 3.9 & 1.2 & 10 \\
4015  & 6000 & 4.0 & 1.3 & 8  \\
5013  & 6050 & 4.0 & 1.3 & 7  \\
5018  & 6050 & 4.2 & 1.3 & 15 \\
6022  & 6020 & 4.0 & 1.3 & 8  \\
7022  & 6000 & 4.0 & 1.4 & 14 \\
12010 & 6050 & 4.2 & 1.3 & 20 \\
12021 & 6000 & 4.3 & 1.2 & 7  \\

\midrule
\multicolumn{5}{l}{\textbf{BSS}}\\
1007  & 7385 & 3.5 & 2.7 & 73 \\
2008  & 5800 & 3.4 & 1.2 & 16 \\
2013  & 7650 & 4.2 & 3.3 & 62 \\
3005  & 7650 & 4.2 & 3.3 & 27 \\
3010  & 6490 & 3.6 & 1.6 & 44 \\
3013  & 7400 & 3.6 & 2.7 & 65 \\

\midrule
\multicolumn{5}{l}{\textbf{YSS}}\\
1015 & 5018 & 2.9 & 1.2 & 7  \\
2002 & 5200 & 3.2 & 1.8 & 14 \\

\midrule
\multicolumn{5}{l}{\textbf{Subgiant}}\\
3014 & 5785 & 3.8  & 1.2 & 11 \\
6008 & 5620 & 3.75 & 1.1 & 8  \\

\midrule
\multicolumn{5}{l}{\textbf{RGB}}\\
1005 & 4300 & 1.7 & 1.6 & 7 \\
2014 & 4680 & 2.5 & 1.3 & 7 \\

\midrule
\multicolumn{5}{l}{\textbf{MS: velocity non-variables}}\\
5017  & 6050 & 4.20 & 1.25 & 8.0 \\
5022  & 5975 & 4.00 & 1.24 & 8.0 \\
7031  & 5900 & 4.30 & 1.50 & 6.0 \\
8012  & 6050 & 4.20 & 1.27 & 5.0 \\
10012 & 6000 & 4.30 & 1.25 & 6.0 \\
10014 & 5800 & 4.30 & 1.50 & 5.0 \\
12025 & 5700 & 4.40 & 1.14 & 5.0 \\

\bottomrule
\end{tabular}
\tablefoot{Evolutionary classifications are based on the CMD in Fig \ref{fig:CMD}.}

\end{table}

\section{Discussion}
\label{sec:discussion}

\subsection{Comments on all notable stars}

WOCS~2013 is a BSS exhibiting RV variations \citep{geller}, and with the highest $s$-process elemental abundances measured in this cluster (this work). WOCS~2013, therefore a Ba dwarf, has a comparatively short orbital period ($P<1000$~d; \citealt{latham1996spectroscopic}) relative to other Case~C post–MT systems in our sample, shows rapid rotation, and displays a UV excess previously identified in UVIT observations by \citet{uocsvi}. Details of the SED fitting and properties of its hot WD companion are provided in \citet{uocsvi}. A complete MT and binary-interaction mechanism of this system should be further explored in the future.
WOCS~3005, which is a BSS (and a Ba dwarf) with enhanced s-process elements, has no detected RV variations \citep{geller}. However, \citet{uocsvi} reported a significant UV excess for this star. The source of this UV excess remains unclear because no convincing hot-companion solution has yet been confirmed from spectroscopy or binary SED modelling, and additional observations are required to understand its origin.

WOCS~2012 and 4015 are Ba dwarfs \citep{escorza2019barium} located near the MS turnoff (Fig \ref{fig:CMD}). While SED analysis confirms the presence of an ELM-WD companion to both these stars, our abundance measurements reveal significant $s$-process enhancements (Table~\ref{tab:abundance}). This presents an evolutionary enigma, as ELM-WD progenitors do not reach the AGB stage, required to synthesise $s$-process elements. These systems may instead be hierarchical triples, as modelled by \citet{lagos2024}, comprising an MS star with an inner ELM WD formed via stable (Case~A or B) Roche lobe overflow and an outer, now faint CO WD donor responsible for the chemical pollution. This configuration is supported by the observed low $v \sin i$ values of 6-8 km$s^{-1}$ and the lack of short-period RV variability from the \citet{geller2021} catalog. Furthermore, our RV variability non-detection analysis (Appendix~\ref{app:rv_limits}) implies long-period motion, with a slow secular trend suggesting that the true period may be substantially longer than the lower limit. Continued long-term RV monitoring is required to confirm this triple-system architecture hypothesis and distinguish it from a simple wide binary configuration.

Hierarchical triple evolution has been increasingly invoked to explain chemically enriched stars in unexpectedly compact binaries (e.g. \citealt{gao2023stellar, lagos2024}). A good example is the Ba dwarf HD~48565, a confirmed triple system in which a Ba-enriched primary resides in a short-period inner binary ($P = 73.344\pm0.013$ d), while the former AGB polluter is now a faint companion on a very long-period orbit ($10 531\pm210$ d; \citealt{escorza2019barium}). One more example is HD~218356, recently identified as a hierarchical triple system with a white dwarf tertiary, where the inner binary ($P = 111.15 \pm 0.03$) consists of a Ba giant and a ELM ($\sim$0.13~$M_\odot$) companion \citep{escorza2023barium}.
Similar triple-origin pathways have been suggested for Ba-enriched binaries and BSSs in OCs (such as WOCS~4004 in NGC~6819, \citealt{milliman2014wiyn, milliman2015barium}), where standard binary evolution alone struggles to reproduce the observed orbital and chemical properties. 
The detailed outcomes of such triple-driven channels likely depend on the specifics of three-body dynamics and mass-transfer efficiency (e.g. \citealt{toonen2016}).
We also present a few similar candidate systems in this study, highlighting the importance of continued searches for tertiary companions in chemically anomalous systems, particularly Ba dwarfs with short orbital periods.

WOCS~12021 is a MS SB1 system with a long, nearly circular orbit ($P=2970\pm70$ d) and clear s-process enhancement. It is a slow rotator and therefore does not meet the observational definition used to identify a blue lurker \citep{leiner2019blue}. Its s-process enhancement and long-period binary orbit are consistent with past Case-C MT from an AGB companion, classifying WOCS~12021 as a Ba dwarf.

WOCS~1007 is a $\delta$ Scuti BSS which shows ellipsoidal variation in a compact ($P\simeq 4.2$~d) binary and has a hot ELM WD companion \citep{geller, uocsi,Vernekar2023MNRAS.524.1360V}. 
Using the orbital period \citep{geller2021} and SED radius of the BSS \citep{uocsi}, we estimate a synchronous rotational velocity of $\sim36$ km s$^{-1}$ ($v_{\rm sync}=2\pi R/P$; \citealt{hilditch2001}), significantly lower than the observed $v\sin i=73$ km s$^{-1}$. This suggests that tidal synchronisation alone cannot account for the rapid rotation and is more consistent with MT-induced spin-up \citep{packet1981}.
However, we detect a modest $s$-process enhancement ($[\mathrm{Y/Fe}] = 0.25 \pm 0.08$; Table~\ref{tab:abundance}), further supporting an interaction-driven history. Given the very short orbital period, a Case~C MT episode is unlikely.
High-resolution spectroscopic studies of $\delta$ Scuti variables \citep{sneden2018} indicate that their photospheres often exhibit overabundances of heavy elements, including s-process species such as Sr, Y, Zr, and Ba. These enhancements are generally attributed to atmospheric processes$-$atomic diffusion, radiative levitation, and rotational effects$-$ rather than active nucleosynthesis \citep{takeda2008,sneden2018, deal2019}. 
The pulsation-related s-process enhancement is much more likely in WOCS~1007 instead of Case~C  MT in a triple system.

An additional pattern can be seen among the ELM WD–MS post–MT binaries in our sample. WOCS~4002 and WOCS~4014 host ELM He-core WD companions, yet RV non-variability constraints imply long present-day orbital period lower limits ($P_{\min}\sim10^{3}$~d; Table~\ref{app:rv_limits}), as derived in the Appendix~\ref{app:rv_limits} section. Both systems show no statistically significant $s$-process enhancement (Table~\ref{tab:abundance}), consistent with Case~A or Case~B MT. Similarly, WOCS~5018 shows no $s$-process enhancement which is consistent with its short period (14.55~d). The coexistence of ELM companions and wide present-day orbital constraints in WOCS~4002 and WOCS~4014 suggests that their current configurations may not directly trace the interaction phase. Similar ELM WD systems in wide binaries have been reported before: for example, HE~0430–2457, which hosts an ELM WD ($M_{\rm WD}\simeq0.23,M_\odot$) in a long-period binary ($P = 771 \pm 3$~d); their binary evolution models suggest formation via merger of an inner binary in a hierarchical triple configuration \citep{vos2018he}. These systems therefore admit multiple pathways, including post–MT orbital modification through cluster dynamics, hierarchical configurations in which the UV/ELM source does not dominate the measured RV baseline, or unfavourable inclination or phase coverage, highlighting the need for continued RV monitoring and searches for a tertiary companion, if any.

\subsection{Orbital characteristics of all systems}

\begin{figure}
    \centering
    \includegraphics[width=1\linewidth]{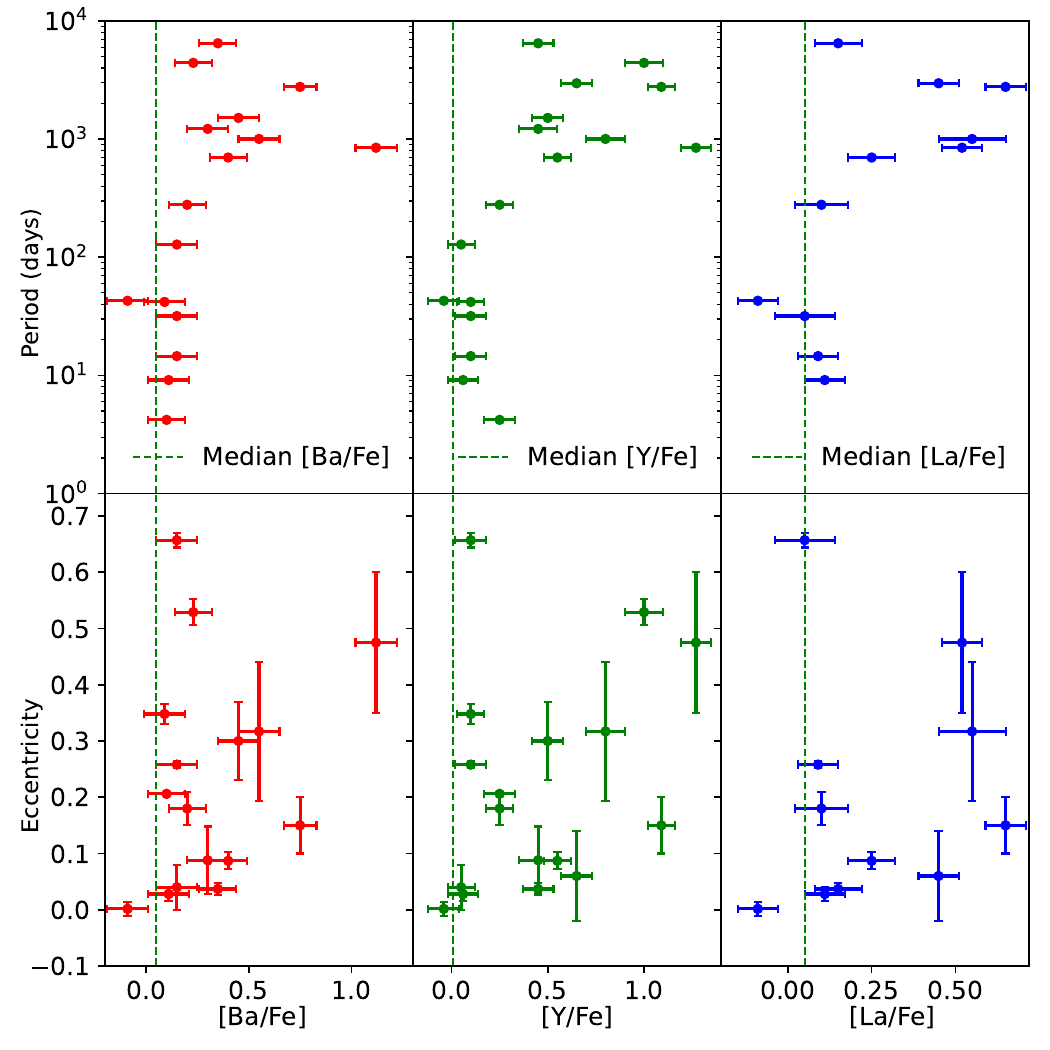}
    \caption{Variation of s-process abundances with period and eccentricity. The median values for the whole cluster are shown by the green dashed line.}
    \label{fig:M67_trend}
\end{figure}

WOCS~5018, with an orbital period of 14.553~days \citep{geller2021}, has a substantially longer period than the compact WD+MS post–CE binaries that dominate the observed population. Its period therefore does not uniquely suggest a CE origin, and is consistent with formation through early, stable RLOF MT (Case~A or Case~B) leading to a He-core/ELM WD as predicted by binary-evolution models \citep{li2019formation}.

For confirmed post-MT binaries without published orbital solutions (WOCS~2012, 4002, 4014, 4015), we use the multi-epoch RVs from \citet{geller2021} to derive lower limits on the orbital period, which are shown in Fig~\ref{fig:Period}. The full computation is shown in the Appendix ~\ref{app:rv_limits} section. These constraints place all of these binaries in the long-period regime, making them particularly valuable targets for future detailed orbital characterisation and modelling to clarify their MT scenarios and formation pathways.

Figure~\ref{fig:M67_trend} shows the distribution of orbital period and eccentricity as a function of s-process abundances ([Ba/Fe], [Y/Fe], [La/Fe]) for all post-MT binaries in M67 in Table~\ref{tab:abundance}. 
A qualitative trend is observed, with the most s-process–enriched systems preferentially found at longer orbital periods, while short-period binaries exhibit little or no enrichment. 
This behaviour mirrors that seen in field MS+WD binaries identified by Gaia, where s-process–enhanced systems predominantly occupy the long-period, moderately eccentric regime \citep{2026ApJ...996L..37R, yamaguchi2025chemical}. 

The similarity between the M67 and field populations indicates that the cluster environment has not fundamentally altered the orbital–chemical imprint of AGB MT; while a small number of systems may deviate from the general trend due to dynamical interactions or subsequent tidal evolution, such effects do not dominate the population. Moreover, the present-day orbital parameters may not uniquely reflect the configuration during mass transfer, as post-MT tidal evolution can substantially modify both orbital periods and eccentricities \citep{Escorza2020A&A...639A..24E}.

\subsection{Comparison with previous studies}

\citet{nine2024wiyn} interpreted Ba enhancement in M67 primarily as evidence for AGB MT in long-period binaries, using Ba as a single s-process tracer. For the most strongly enriched blue stragglers (e.g. WOCS~2013 and WOCS~3005), our measured Ba abundances are consistent with the ranges reported by \citet{nine2024wiyn}. However, for Yellow stragglers (WOCS~1015 and 2008) and blue lurkers (WOCS~9005), \citet{nine2024wiyn} report no evidence of Ba enrichment, which they interpret as expected if the dominant formation channel is mass transfer from an RGB donor, in Case~B MT scenario prior to the helium flash. In contrast, we measure  higher [Ba/Fe] (and Y and La), by up to $\sim$0.4--0.7\, dex for these stars. This difference for WOCS~9005 has been discussed in \citet{pal2024discovery}, where the sensitivity of the Ba abundance to the adopted microturbulent velocity is shown to account for the difference with \citet{nine2024wiyn}. The presence of clear s-process enrichment in these systems is therefore explained by Case~C MT from an AGB donor rather than Case~B RGB MT.

\citet{mota} carried out a spectroscopic study of M67 BSS and TO cluster members, using mainly C and N abundances from high-resolution spectroscopy (APOGEE) to search for chemical signatures of post–MT evolution, and found that the abundance patterns are largely consistent with normal stellar evolution (e.g. post–first dredge-up mixing) rather than clear MT effects. For the stars common to our sample: WOCS~2012 (S756), 4015 (S1456), 2013 (S1267), and 3010 (S975), they treated these objects as turn-off or BSS stars and did not report abundance patterns that would classify them as post–MT binaries. In contrast, we find strong enrichment of s-process elements in these same stars (Table~\ref{tab:abundance}), indicating that they have undergone MT. This conclusion is further supported by other studies showing that carbon abundances are a poor diagnostic of mass transfer due to NLTE uncertainties and the small expected abundance changes, as demonstrated by \citep{brady2023m67} with a detailed analysis of M67 stars like WOCS~3010.

\section{Summary and conclusion}
\label{sec:summary}

This paper presents our study on M67 using FUV \textit{AstroSat}/UVIT photometry \citep{uocsiii}, NUV data from \textit{Swift}/UVOT \citep{siegel3} and other archival optical/IR photometry. Among a total of 141 UV-detected cluster members, we excluded ten stars from the SED fitting due to crowding and higher-order multiplicity, resulting in a final dataset of 132 stars. Among them, 22 sources showed UV excess and were fitted with binary models.

The analysis of the binary SEDs showed five previously undetected hot companions in MS+ ELM WD binary systems which have undergone MT in past, representing a significant finding for the cluster. We estimated the photometric masses of WD components using WD cooling curves (Table~\ref{table:dynamic}). 
Other than these 5 confirmed ELM WDs and MS binaries, we also present 6 previously un-detected hot companions in MS+WD binary systems (but with uncertain WD types; Appendix ~\ref{app:UV_excess}).
There was no relation found between the WD mass and the stellar rotation (a potential indicator of MT  epoch) in any of these systems. The number density of the 5 ELM WDs present in M67 is orders of magnitude higher than that of the Galactic field \citep{li2019formation}. 

We performed a spectral analysis of the stars in M67 using GALAH \citep{galah} surveys and compared the elemental abundances to check for any post-MT abundance variations (summarised in Table~\ref{tab:abundance}). Stars with longer orbital periods (>1 year) predominantly show enhancement of s-process elements. These enhancements in M67 are consistent with Case~C MT from thermally pulsing AGB companions that have since evolved into optically faint WDs, and these chemically enriched post-MT systems span multiple evolutionary stages (MS, BSS, YSS, subgiant, and red giant), indicating that Case~C MT signatures can be seen across the cluster CMD. Following earlier evidence for AGB MT as a major formation channel for BSSs in old open clusters \citep{Gosnell2015, nine2024wiyn, mathieu2025blue}, we find s-process enhancement in $\gtrsim33\%$ of the studied BSSs in M67, including some non–RV-variable stars (e.g. WOCS~3005).

Overall, this multiwavelength analysis reports that M67 has five new ELM WD-MS binary systems (and six WD-MS systems) which were formed via MT; 13 s-process enhanced binary systems which were likely formed via Case~C MT from an AGB star. These post-MT systems lie on the various
stages of stellar evolution: MS, BSS, YSS, subgiant and giant branch. The UV-IR multiwavelength study highlights the diversity among the post-MT systems within M67. Our study combines three elements-s-process abundances (Ba, Y, La) with UV-based SED analysis, enabling us to distinguish between Case~A or B  and Case~C MT channels. This unified framework demonstrates that all classical MT pathways operate concurrently in M67 at every evolutionary stage, providing a more complete picture of binary evolution within the cluster. 
There is a large sample of diverse post-MT systems to be found in open clusters. Future multi-wavelength surveys and long-baseline spectroscopic observations will be crucial in understanding the frequency of such MT phenomena and their effect on stellar evolution and parameterization.

\begin{acknowledgements}
    We thank the anonymous referee for constructive comments that helped improve the quality of this manuscript. 
    AS thanks for the support of the SERB power fellowship. VJ acknowledges support from the Alexander von Humboldt Foundation and grant 26-21774S from the Czech Grant Agency. This work uses the data from UVIT onboard the \textit{AstroSat} mission of the Indian Space Research Organisation (ISRO). UVIT is a collaborative project involving various institutes, including the Indian Institute of Astrophysics (IIA), Bengaluru, the Inter-University Center for Astronomy and Astrophysics (IUCAA), Pune, the Tata Institute of Fundamental Research (TIFR), Mumbai, several centerthe s of the Indian Space Research Organization (ISRO), and the Canadian Space Agency (CSA). This publication uses VOSA, developed under the Spanish Virtual Observatory project supported by the Spanish MINECO through grant AyA2017-84089. 
    For this research, we also utilised several tools and databases, including the VizieR catalogue access tool provided by CDS in Strasbourg, France, TOPCAT \citep{taylor2005topcat}, Python modules like Matplotlib \citep{hunter2007matplotlib}, NumPy \citep{harris2020Numpy}, pandas \citep{reback2020pandas} and NASA's Astrophysics Data System (ADS NASA).
\end{acknowledgements}

\textbf{Data availability.}
The catalogue of this study (the full version of Table~\ref{tab:SED}, combined with the data of Tables~\ref{table:dynamic}, \ref{tab:params}, \ref{table:wd_candidates}, \ref{tab:abundance}, and \ref{tab:rv_pmin}) is available in electronic form at the CDS via anonymous ftp to \url{cdsarc.cds.unistra.fr} (130.79.128.5) or via \url{https://cdsarc.cds.unistra.fr/cgi-bin/qcat?J/A+A/}.
It extends Table~\ref{tab:SED} to all 152 stars studied and combines the information from Tables~\ref{table:dynamic}, \ref{tab:params}, \ref{table:wd_candidates}, \ref{tab:abundance}, and \ref{tab:rv_pmin}.
For each star, the catalogue provides the AstroSat/UVIT, Swift/UVOT, and Gaia photometry; the SED-fit parameters; the WD masses and cooling ages; the orbital parameters and RV-based period limits; the spectroscopic stellar parameters; the Fe, Y, Ba, and La abundances; and our MT classification.

\bibliographystyle{aa_url} 
\bibliography{example}

\begin{appendix}
\nolinenumbers

\section{UV-excess detected stars: MS+WD binary systems}\label{app:UV_excess}

\begin{sidewaysfigure*}
\centering
\includegraphics[width=1\textwidth]{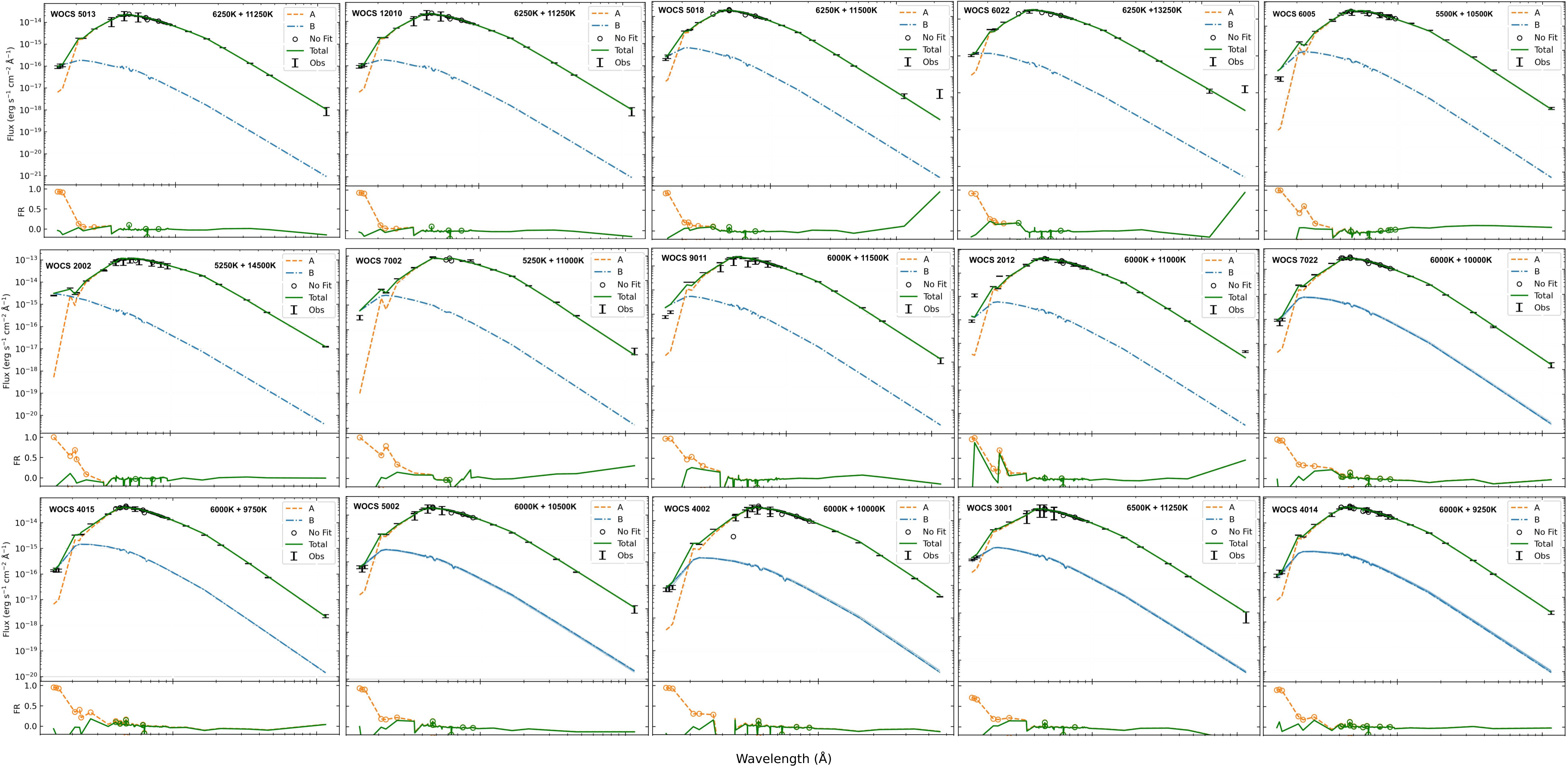}
\caption{SEDs of 15 stars exhibiting significant UV excesses. The top two rows display ten stars fitted using Kurucz (orange dot-dashed) and Koester WD models (blue dashed), while the bottom row shows five stars fitted with a hybrid Kurucz+UVBlue (orange dashed) and Koester combination. For each SED, the panel shows the observed fluxes (black error bars), excluded data points (shown as open circles), individual model components (orange and blue), and the combined model flux (green solid line). The bottom panel shows fractional residuals for single-component (orange dashed) and two-component (green solid) fits. Fractional observational uncertainties are indicated along the wavelength axis. }
    \label{fig:SED}
\end{sidewaysfigure*}

\begin{table}\centering
\caption{The MS+WD binary candidates with successful SED binary fits but unconfirmed WD types.}
\small
\renewcommand{\arraystretch}{1.3}
\setlength{\tabcolsep}{3pt}
\begin{tabular}{rccccc}
\toprule
WOCS & Stage & M$_{WD}$ ($\mathrm{M}_{\odot}$) & WD Type & log(age/yr) & X-ray \\
     &       & (Mean$\pm$errors)              &          & (Mean$\pm$errors) & Emission\\
\midrule
5013  & MS   & $0.176_{-0.026}^{+0.824}$  & ELM WD$^{\ddagger}$   & $9.249_{-0.830}^{+0.617}$ & No\\
12010 & MS   & $0.176_{-0.026}^{+0.824}$  & ELM WD$^{\ddagger}$   & $9.256_{-0.837}^{+0.773}$ & No\\
6022  & MS   & $0.2_{-0.05}^{+1.0}$       & ELM WD$^{\ddagger}$   & $7.042_{-1.478}^{+2.737}$ & No\\
6005  & SGB  & $0.161_{-0.011}^{+0.019}$  & ELM WD                 & $9.208_{-0.724}^{+0.172}$ & Yes\\
2002  & YSS  & $0.182_{-0.002}^{+0.008}$  & ELM WD                 & $9.037_{-0.233}^{+0.002}$ & Yes\\
7002  & SSG  & $0.2_{-0.03}^{+0.3}$       & ELM WD$^{\ddagger}$   & $6.840_{-0.611}^{+2.439}$ & Yes\\
9011  & MS   & $0.182_{-0.027}^{+0.020}$ & ELM WD                 & $8.977_{-0.558}^{+0.488}$ & Yes\\
7022  & MS   & $0.5_{-0.35}^{+0.8}$       & WD$^{\ddagger}$        & $8.346_{-2.782}^{+1.193}$ & No\\
5002  & MS   & $0.182_{-0.032}^{+0.618}$ & ELM WD$^{\ddagger}$   & $8.959_{-0.540}^{+0.979}$ & No\\
3001  & MS   & $0.165_{-0.010}^{+0.835}$ & ELM WD$^{\ddagger}$   & $9.138_{-0.719}^{+0.665}$ & No\\
\bottomrule
\end{tabular}
\tablefoot{Approximate WD masses and cooling ages are estimated using WD cooling curves by \citet{WD} models (for $0.2-1.3 \mathrm{M}_{\odot}$) and \citet{elm} models (for $< 0.2 \mathrm{M}_{\odot}$). Stellar evolutionary stages are determined from CMDs in Figure~\ref{fig:CMD}.
$^{\ddagger}$ Indicates WD mass ranges whose error bars extend across multiple WD-type regimes.
}
\label{table:wd_candidates}
\end{table}

There are six UV-excess stars (without X-ray detections) that are well fit by MS+WD binary SED models but lack sufficiently constrained WD masses for firm classification. Spectroscopic orbits for 3 UV-excess binaries: WOCS~3001, WOCS~12010, and WOCS~7022, show periods of $\sim$9–128~days and eccentricities of 0.02–0.35, consistent with systems that have experienced  stable RLOF MT \citep{geller2021}. 
\citet{nine2023wiyn} used HST far-UV photometry and identified a significant FUV excess in WOCS~3001. After analysis, it was noted that to be a Case~B post-MT system.

In the four Xray detected sources, WOCS~6005 (S999/HW~Cnc), 7002 (S1019), and 9011 (S773) are all short-period ($P \lesssim 10$~d), nearly circular, X-ray--bright binaries in M67 whose properties are consistent with chromospherically active RS~CVn (RS~Canum~Venaticorum) type systems \citep{van2002photometric,Van,geller2021}. \citet{belloni1998} classify WOCS~6005 (S999/HW~Cnc) as an RS~CVn system and argue that WOCS~7002 (S1019) is also an RS~CVn candidate based on its ROSAT luminosity and short orbital period, adopting a definition that encompasses BY~Draconis--type active MS binaries, while \citet{van2002photometric} show that the X-ray emission of WOCS~6005 (S999) arises from tidally locked magnetically active components. \citet{geller2021} further confirm the short-period SB2 nature of these systems. We therefore do not interpret the detected UV excess in these systems as arising from MS+WD or ELM companions (from the SED fitting) and also exclude them from Fig~\ref{fig:Period}. 

However, despite its X-ray detection, WOCS~2002 (S1040) is a well-established YSS + He-core WD system. Its UV excess was first detected with UIT imaging of M67, and the hot WD companion was later confirmed with HST/GHRS far-UV spectroscopy. It has been interpreted as an evolved blue straggler formed through early Case~B MT and binary evolution, leaving a low-mass WD companion \citep{landsman1997s1040,mota}; we therefore include this system in Fig~\ref{fig:Period} as well to compare its orbital period with those of other post-MT binaries.

WOCS~5013 and 6022 are fundamentally similar MS stars. Both lie at almost the same location in the M67 CMD (Fig~\ref{fig:CMD}) and share similar primary mass estimates (Table~\ref{table:dynamic}) and $T_{eff}$ (Table~\ref {tab:SED}), all stellar parameters (Table~\ref {tab:params}) and both exhibit clear UV excesses indicative of hot WD companions. The SED fitting indicates WD masses of 0.15–1.0~$M_\odot$ for WOCS~5013 and 0.15–1.2~$M_\odot$ for WOCS~6022 (Table~\ref{table:wd_candidates}). From a chemical perspective, though, WOCS~6022 shows evident $s$-process enhancement (SEMS; Table~\ref{tab:abundance}), whereas WOCS~5013 does not. These systems may therefore have formed through different post-MT evolutionary pathways despite their similar stellar parameters and CMD locations.
The available WIYN/CfA RV time series for WOCS~5013 and 6022 show only $\lesssim$1 km s$^{-1}$ peak-to-peak excursions, comparable to the $\sim$0.5 km s$^{-1}$ WIYN precision, and therefore do not meet the survey’s variability threshold for secure orbital characterisation \citep{geller2021}. Continued spectroscopic follow-up and RV monitoring are essential to understand their formation mechanisms.

\begin{table*}
\centering
\begin{threeparttable}
\caption{Elemental abundances ([X/Fe]) with metallicity ([Fe/H]) for a sample of confirmed binary members of M67.}
\label{tab:heavy_elements_galah_vs_thisstudy}
\fontsize{8.2}{8.8}\selectfont
\setlength{\tabcolsep}{4pt}

\begin{tabular}{lllcc lccc cccc}
\toprule
WOCS & Stage & Class & $P$ & $e$ & Source & [Fe/H] & [Y/Fe] & [Ba/Fe] & [La/Fe] \\
\midrule
\multicolumn{10}{l}{\textbf{Case~C post-MT binaries}}\\
\midrule

\multirow{2}{*}{2012} & \multirow{2}{*}{MS} & \multirow{2}{*}{SM} & \multirow{2}{*}{$\ldots$} & \multirow{2}{*}{$\ldots$}
& GALAH     & 0.00$\pm$0.05  & 0.09$\pm$0.05  & 0.21$\pm$0.04  & $\ldots$ \\
& & & & 
& This study & -0.05$\pm$0.05 & 0.40$\pm$0.08  & 0.45$\pm$0.09  & 0.20$\pm$0.06 \\

\multirow{2}{*}{4015} & \multirow{2}{*}{MS} & \multirow{2}{*}{SM} & \multirow{2}{*}{$\ldots$} & \multirow{2}{*}{$\ldots$}
& GALAH     & -0.08$\pm$0.05 & 0.22$\pm$0.07  & 0.08$\pm$0.05  & 0.29$\pm$0.08 \\
& & & & 
& This study & -0.10$\pm$0.05 & 0.35$\pm$0.08  & 0.30$\pm$0.10  & 0.22$\pm$0.06 \\

\multirow{2}{*}{12021} & \multirow{2}{*}{MS} & \multirow{2}{*}{BM, SB1} & \multirow{2}{*}{2970$\pm$70} & \multirow{2}{*}{0.06$\pm$0.08}
& GALAH     & -0.02$\pm$0.05 & 0.35$\pm$0.05  & -0.06$\pm$0.04 & 0.63$\pm$0.07 \\
& & & & 
& This study & -0.05$\pm$0.05 & 0.65$\pm$0.08  & $0.35\pm0.08$       & 0.45$\pm$0.06 \\

\multirow{2}{*}{2008} & \multirow{2}{*}{BSS} & \multirow{2}{*}{BM, SB1} & \multirow{2}{*}{1513$\pm$7} & \multirow{2}{*}{0.3$\pm$0.07}
& GALAH     & -0.10$\pm$0.05 & 0.15$\pm$0.09  & 0.18$\pm$0.05  & -0.01$\pm$0.08 \\
& & & & 
& This study & -0.10$\pm$0.05 & 0.50$\pm$0.08  & 0.45$\pm$0.10  & $\ldots$ \\

\multirow{2}{*}{2013} & \multirow{2}{*}{BSS} & \multirow{2}{*}{BM, SB1} & \multirow{2}{*}{846 [L96]} & \multirow{2}{*}{0.475$\pm$0.125[L96]}
& GALAH     & -0.41$\pm$0.06 & 0.79$\pm$0.06  & 0.68$\pm$0.04  & $\ldots$ \\
& & & & 
& This study & -0.07$\pm$0.04 & 1.27$\pm$0.08  & 1.12$\pm$0.10  & 0.52$\pm$0.06 \\

\multirow{2}{*}{3005} & \multirow{2}{*}{BSS} & \multirow{2}{*}{SM} & \multirow{2}{*}{No data} & \multirow{2}{*}{No data}
& GALAH     & +0.04$\pm$0.05 & 0.75$\pm$0.08  & 0.66$\pm$0.06  & $\ldots$ \\
& & & & 
& This study & -0.07$\pm$0.04 & 1.06$\pm$0.08  & 0.94$\pm$0.10  & 0.51$\pm$0.06 \\

\multirow{2}{*}{3010} & \multirow{2}{*}{BSS} & \multirow{2}{*}{BM, SB1} & \multirow{2}{*}{1221 [L96]} & \multirow{2}{*}{0.088$\pm$0.06[L96]}
& GALAH     & -0.15$\pm$0.05 & 0.62$\pm$0.05  & 0.16$\pm$0.04  & $\ldots$ \\
& & & & 
& This study & -0.10$\pm$0.05 & 0.45$\pm$0.10  & 0.30$\pm$0.10  & $\ldots$ \\

\multirow{2}{*}{3013} & \multirow{2}{*}{BSS} & \multirow{2}{*}{BM, SB1} & \multirow{2}{*}{1003 [L96]} & \multirow{2}{*}{0.317$\pm$0.123[L96]}
& GALAH     & -0.57$\pm$0.05 & 0.77$\pm$0.05  & $\ldots$       & $\ldots$ \\
& & & & 
& This study & -0.05$\pm$0.05 & 0.80$\pm$0.10  & 0.55$\pm$0.10  & 0.55$\pm$0.10 \\

\multirow{2}{*}{9005} & \multirow{2}{*}{BSS} & \multirow{2}{*}{BM, SB1} & \multirow{2}{*}{2769$\pm$18} & \multirow{2}{*}{0.15$\pm$0.05}
& GALAH     & -0.07$\pm$0.06 & 1.11$\pm$0.07  & 0.64$\pm$0.05  & 0.68$\pm$0.07 \\
& & & & 
& Pal2024 & -0.10$\pm$0.05 & 1.09$\pm$0.07  & 0.75$\pm$0.08  & 0.65$\pm$0.06 \\

\multirow{2}{*}{1015} & \multirow{2}{*}{YSS} & \multirow{2}{*}{BM, SB1} & \multirow{2}{*}{698.56$\pm$0.24} & \multirow{2}{*}{0.087$\pm$0.015}
& GALAH     & -0.01$\pm$0.04 & 0.26$\pm$0.05  & 0.33$\pm$0.03  & -0.12$\pm$0.01 \\
& & & & 
& This study & -0.10$\pm$0.05 & 0.55$\pm$0.07  & 0.40$\pm$0.09  & 0.25$\pm$0.07 \\

\multirow{2}{*}{3014} & \multirow{2}{*}{Subgiant} & \multirow{2}{*}{BM,SB1} & \multirow{2}{*}{278.2$\pm$1.1} & \multirow{2}{*}{0.18$\pm$0.03}
& GALAH     & -0.12$\pm$0.05 & 0.13$\pm$0.06  & 0.07$\pm$0.04  & $\ldots$ \\
& & & & 
& This study & -0.10$\pm$0.05 & 0.25$\pm$0.07  & 0.20$\pm$0.09  & 0.10$\pm$0.08 \\

\multirow{2}{*}{1005} & \multirow{2}{*}{Red Giant} & \multirow{2}{*}{BM,SB1} & \multirow{2}{*}{4422$\pm$17} & \multirow{2}{*}{0.529$\pm$0.023}
& GALAH     & -0.28$\pm$0.04 & 0.23$\pm$0.06  & 0.26$\pm$0.04  & 0.04$\pm$0.01 \\
& & & & 
& This study & -0.08$\pm$0.05 & 1.00$\pm$0.10  & 0.23$\pm$0.09  & $\ldots$ \\

\multirow{2}{*}{2014} & \multirow{2}{*}{Red Giant} & \multirow{2}{*}{BM,SB1} & \multirow{2}{*}{6475$\pm$16} & \multirow{2}{*}{0.037$\pm$0.011}
& GALAH     & +0.03$\pm$0.04 & -0.26$\pm$0.05 & 0.21$\pm$0.04  & -0.05$\pm$0.03 \\
& & & & 
& This study & +0.05$\pm$0.05 & 0.45$\pm$0.08  & 0.35$\pm$0.09  & 0.15$\pm$0.07 \\

\midrule
\multicolumn{10}{l}{\textbf{Confirmed Case~A or Case~B post-MT binaries}}\\
\midrule

\multirow{2}{*}{4002} & \multirow{2}{*}{MS} & \multirow{2}{*}{SM} & \multirow{2}{*}{$\ldots$} & \multirow{2}{*}{$\ldots$}
& GALAH     & -0.08$\pm$0.09 & 0.09$\pm$0.18  & 0.04$\pm$0.12  & $\ldots$ \\
& & & & 
& This study & -0.09$\pm$0.05 & 0.10$\pm$0.08  & $0.03\pm0.09$       & 0.09$\pm$0.08 \\

\multirow{2}{*}{4014} & \multirow{2}{*}{MS} & \multirow{2}{*}{SM} & \multirow{2}{*}{$\ldots$} & \multirow{2}{*}{$\ldots$}
& GALAH     & -0.05$\pm$0.06 & 0.27$\pm$0.09  & $\ldots$       & $\ldots$ \\
& & & & 
& This study & 0.00$\pm$0.05  & $\ldots$       & 0.10$\pm$0.10  & 0.00$\pm$0.07 \\

\multirow{2}{*}{5018} & \multirow{2}{*}{MS} & \multirow{2}{*}{BM, SB1} & \multirow{2}{*}{14.5533$\pm$0.0001} & \multirow{2}{*}{0.258$\pm$0.006}
& GALAH     & -0.14$\pm$0.06 & -0.11$\pm$0.07 & -0.02$\pm$0.05 & $\ldots$ \\
& & & & 
& This study & -0.09$\pm$0.05 & 0.10$\pm$0.08  & 0.15$\pm$0.10  & 0.09$\pm$0.06 \\

\multirow{2}{*}{1007} & \multirow{2}{*}{BSS} & \multirow{2}{*}{BM, SB1} & \multirow{2}{*}{4.2$\pm$0.041[V23]} & \multirow{2}{*}{0.206$\pm$0.002[V23]}
& GALAH     & -0.82$\pm$0.07 & 0.39$\pm$0.08  & $\ldots$       & $\ldots$ \\
& & & & 
& This study & -0.10$\pm$0.05 & 0.25$\pm$0.08  & 0.10$\pm$0.09  & $\ldots$ \\

\midrule
\multicolumn{10}{l}{\textbf{Candidates of Case~A or Case~B post-MT binaries (UV excess and close period Binaries)}}\\
\midrule

\multirow{2}{*}{3001} & \multirow{2}{*}{MS} & \multirow{2}{*}{BM, SB1} & \multirow{2}{*}{128.14$\pm$0.03} & \multirow{2}{*}{0.04$\pm$0.04}
& GALAH     & -0.08$\pm$0.09 & 0.06$\pm$0.12  & -0.16$\pm$0.09 & $\ldots$ \\
& & & & 
& This study & -0.05$\pm$0.05 & 0.05$\pm$0.07  & 0.15$\pm$0.10  & $\ldots$ \\

\multirow{2}{*}{7022} & \multirow{2}{*}{MS} & \multirow{2}{*}{BM, SB1} & \multirow{2}{*}{9.12707$\pm$0.0001} & \multirow{2}{*}{0.028$\pm$0.013}
& GALAH     & -0.08$\pm$0.05 & 0.06$\pm$0.05  & 0.16$\pm$0.04  & $\ldots$ \\
& & & & 
& This study & -0.12$\pm$0.05 & 0.06$\pm$0.08  & 0.11$\pm$0.10  & 0.11$\pm$0.06 \\

\multirow{2}{*}{12010} & \multirow{2}{*}{MS} & \multirow{2}{*}{BM, SB1} & \multirow{2}{*}{41.868$\pm$0.024} & \multirow{2}{*}{0.348$\pm$0.017}
& GALAH     & -0.08$\pm$0.05 & -0.06$\pm$0.06 & 0.03$\pm$0.04  & $\ldots$ \\
& & & & 
& This study & -0.09$\pm$0.05 & 0.10$\pm$0.07  & 0.09$\pm$0.10  & $\ldots$ \\

\multirow{2}{*}{2002} & \multirow{2}{*}{YSS} & \multirow{2}{*}{BM, SB1} & \multirow{2}{*}{42.8251$\pm$0.001} & \multirow{2}{*}{0.0016$\pm$0.012}
& GALAH     & -0.06$\pm$0.04 & 0.21$\pm$0.05  & 0.35$\pm$0.04  & 0.37$\pm$0.09 \\
& & & & 
& This study & -0.12$\pm$0.04 & -0.04$\pm$0.08 & -0.09$\pm$0.10 & -0.09$\pm$0.06 \\

\multirow{2}{*}{6008} & \multirow{2}{*}{Subgiant} & \multirow{2}{*}{BM,SB1} & \multirow{2}{*}{31.7774$\pm$0.0006} & \multirow{2}{*}{0.657$\pm$0.013}
& GALAH     & -0.08$\pm$0.04 & 0.15$\pm$0.05  & 0.20$\pm$0.04  & 0.17$\pm$0.08 \\
& & & & 
& This study & -0.10$\pm$0.05 & 0.10$\pm$0.08  & 0.15$\pm$0.10  & 0.05$\pm$0.09 \\

\midrule
\multicolumn{10}{l}{\textbf{UV Excess}}\\
\midrule

\multirow{2}{*}{5013} & \multirow{2}{*}{MS} & \multirow{2}{*}{SM} & \multirow{2}{*}{$\ldots$} & \multirow{2}{*}{$\ldots$}
& GALAH     & -0.09$\pm$0.05 & 0.08$\pm$0.07  & 0.04$\pm$0.05  & 0.29$\pm$0.11 \\
& & & & 
& This study & -0.05$\pm$0.05 & 0.10$\pm$0.08  & 0.13$\pm$0.09  & 0.05$\pm$0.06 \\

\multirow{2}{*}{6022} & \multirow{2}{*}{MS} & \multirow{2}{*}{SM} & \multirow{2}{*}{$\ldots$} & \multirow{2}{*}{$\ldots$}
& GALAH     & -0.08$\pm$0.05 & -0.09$\pm$0.05 & -0.05$\pm$0.04 & $\ldots$ \\
& & & & 
& This study & -0.10$\pm$0.05 & 0.25$\pm$0.08  & 0.40$\pm$0.10  & 0.25$\pm$0.07 \\

\midrule
\multicolumn{10}{l}{\textbf{MS: non-velocity variables}}\\
\midrule

\multirow{2}{*}{5017}  & \multirow{2}{*}{MS} & \multirow{2}{*}{$\ldots$} & \multirow{2}{*}{$\ldots$} & \multirow{2}{*}{$\ldots$}
& GALAH     & -0.06$\pm$0.07 & -0.31$\pm$0.10 & 0.10$\pm$0.07 & 0.46$\pm$0.14 \\
& & & & 
& This study & -0.10$\pm$0.05 & 0.05$\pm$0.06  & 0.05$\pm$0.08 & 0.05$\pm$0.07 \\

\multirow{2}{*}{5022}  & \multirow{2}{*}{MS} & \multirow{2}{*}{$\ldots$} & \multirow{2}{*}{$\ldots$} & \multirow{2}{*}{$\ldots$}
& GALAH     & -0.15$\pm$0.06 & 0.31$\pm$0.08  & -0.04$\pm$0.06 & 0.46$\pm$0.14 \\
& & & & 
& This study & -0.05$\pm$0.05 & 0.15$\pm$0.08  & 0.06$\pm$0.08  & 0.02$\pm$0.07 \\

\multirow{2}{*}{7031}  & \multirow{2}{*}{MS} & \multirow{2}{*}{$\ldots$} & \multirow{2}{*}{$\ldots$} & \multirow{2}{*}{$\ldots$}
& GALAH     & -0.04$\pm$0.12 & 0.10$\pm$0.37  & -0.06$\pm$0.17 & $\ldots$ \\
& & & & 
& This study & 0.00$\pm$0.05  & 0.00$\pm$0.06  & -0.05$\pm$0.09 & $\ldots$ \\

\multirow{2}{*}{8012}  & \multirow{2}{*}{MS} & \multirow{2}{*}{$\ldots$} & \multirow{2}{*}{$\ldots$} & \multirow{2}{*}{$\ldots$}
& GALAH     & -0.07$\pm$0.05 & 0.02$\pm$0.07  & 0.00$\pm$0.05  & $\ldots$ \\
& & & & 
& This study & -0.09$\pm$0.05 & 0.09$\pm$0.07  & 0.04$\pm$0.08  & -0.01$\pm$0.07 \\

\multirow{2}{*}{10012} & \multirow{2}{*}{MS} & \multirow{2}{*}{$\ldots$} & \multirow{2}{*}{$\ldots$} & \multirow{2}{*}{$\ldots$}
& GALAH     & -0.10$\pm$0.06 & -0.09$\pm$0.11 & -0.04$\pm$0.07 & $\ldots$ \\
& & & & 
& This study & -0.05$\pm$0.05 & 0.05$\pm$0.06  & 0.00$\pm$0.10  & -0.05$\pm$0.08 \\

\multirow{2}{*}{10014} & \multirow{2}{*}{MS} & \multirow{2}{*}{$\ldots$} & \multirow{2}{*}{$\ldots$} & \multirow{2}{*}{$\ldots$}
& GALAH     & -0.13$\pm$0.14 & $\ldots$   & 0.08$\pm$0.17  & $\ldots$ \\
& & & & 
& This study & -0.09$\pm$0.05 & 0.09$\pm$0.06  & 0.12$\pm$0.09  & $\ldots$ \\

\multirow{2}{*}{12025} & \multirow{2}{*}{MS} & \multirow{2}{*}{$\ldots$} & \multirow{2}{*}{$\ldots$} & \multirow{2}{*}{$\ldots$}
& GALAH     & -0.04$\pm$0.05 & 0.14$\pm$0.06  & 0.01$\pm$0.04  & 0.41$\pm$0.09 \\
& & & & 
& This study & 0.00$\pm$0.05  & 0.00$\pm$0.06  & 0.10$\pm$0.09  & $\ldots$ \\
\bottomrule
\end{tabular}
\label{tab:abundance}
\tablefoot{For each system, the first row lists GALAH DR3 automated pipeline abundances \citep{galah} and the second row lists abundances measured in this study. Orbital parameters (inc. Class) are primarily from \citet{geller2021}, while values marked [L96], [V23], and Pal2024 are from \citet{latham1996spectroscopic}, \citet{Vernekar2023MNRAS.524.1360V}, and \citet{pal2024discovery}, respectively. "$\ldots$" denotes missing entries from the source. }
\end{threeparttable}
\end{table*}

\section{RV non-variability and period lower limits}
\label{app:rv_limits}

For confirmed ELM WD + MS post-MT binary systems without complete spectroscopic orbit solutions (WOCS~2012, 4002, 4014, 4015), we derive minimum orbital periods from the RV variability non-detection in the time-series data of \citet{geller2021}. For each star, we compute the time baseline $\Delta T$ and the rms scatter about the mean RV, $\sigma_{\rm RV}$, and adopt a conservative non-detection threshold $K_{\rm lim}=3\sigma_{\rm RV}$, consistent with typical RV detectability requirements of $K \approx 2$--$4\,\sigma$ for time-series with $N\sim10$--20 observations \citep{cumming2004}. Any coherent Keplerian signal with $K \ge K_{\rm lim}$ would be expected to produce detectable variability in the available time series. The RV semi-amplitude for an SB1 is
\begin{equation}
K = \left(\frac{2\pi G}{P}\right)^{1/3}
\frac{M_2 \sin i}{(M_1+M_2)^{2/3}}
\frac{1}{\sqrt{1-e^2}} \, ,
\end{equation}
so setting $K=K_{\rm lim}$ yields the minimum period consistent with the non-detection:
\begin{equation}
P_{\min} =
\frac{2\pi G}{K_{\rm lim}^3}
\frac{M_2^3 \sin^3 i}{(M_1+M_2)^2}
(1-e^2)^{-3/2}.
\end{equation}

We treat $\sin i$ statistically and assume randomly oriented orbital axes, for which the inclination distribution is $p(i)\,di = \sin i\,di$ \citep[e.g.][]{grether2006}. Because Eq.~(B.1) scales as $\sin^3 i$, the relevant ensemble factor is $\langle \sin^3 i\rangle$; for isotropic orientations one obtains $\langle \sin^3 i\rangle = 3\pi/16$, implying an effective $\sin i_{\rm eff}\equiv \langle \sin^3 i\rangle^{1/3}\simeq 0.84$ \citep[e.g.][]{grether2006}. We adopt the WD masses $M_2$ from Table~\ref{table:dynamic}, primary masses $M_1$ from the isochrone fits (Table~\ref {table:dynamic}), and assume $e=0$, which yields the least restrictive period limit for a given $K_{\rm lim}$.
The resulting period lower limits $P_{\min}$ are listed in Table~\ref{tab:rv_pmin}.

\begin{table}[ht]
\centering
\caption{RV non-variability constraints and period lower limits for UV-excess systems without published orbital solutions.}
\label{tab:rv_pmin}
\begin{tabular}{lccc}
\hline
WOCS & $N_{\rm RV}$ & $\sigma_{\rm RV}$ (km s$^{-1}$) & $P_{\min}$ (d) \\
\hline
2012 & 12 & 0.689 & $1.46\times 10^{3}$ \\
4002 &  7 & 0.532 & $3.45\times 10^{3}$ \\
4014 & 10 & 0.675 & $1.96\times 10^{3}$ \\
4015 & 25 & 1.373 & $2.76\times 10^{2}$ \\
\hline
\end{tabular}
\tablefoot{We list the number of RV epochs $N_{\rm RV}$, the rms scatter about the mean $\sigma_{\rm RV}$, and the corresponding period lower limit $P_{\min}$. 
For WOCS~4015, the formal non-variability limit is $P_{\min}\simeq 2.8\times10^{2}\,$d; however, the slow secular trend in the RVs suggests a period of $\gtrsim4\times10^{3}\,$d.
}
\end{table}

\begin{figure}
\includegraphics[width=\columnwidth]{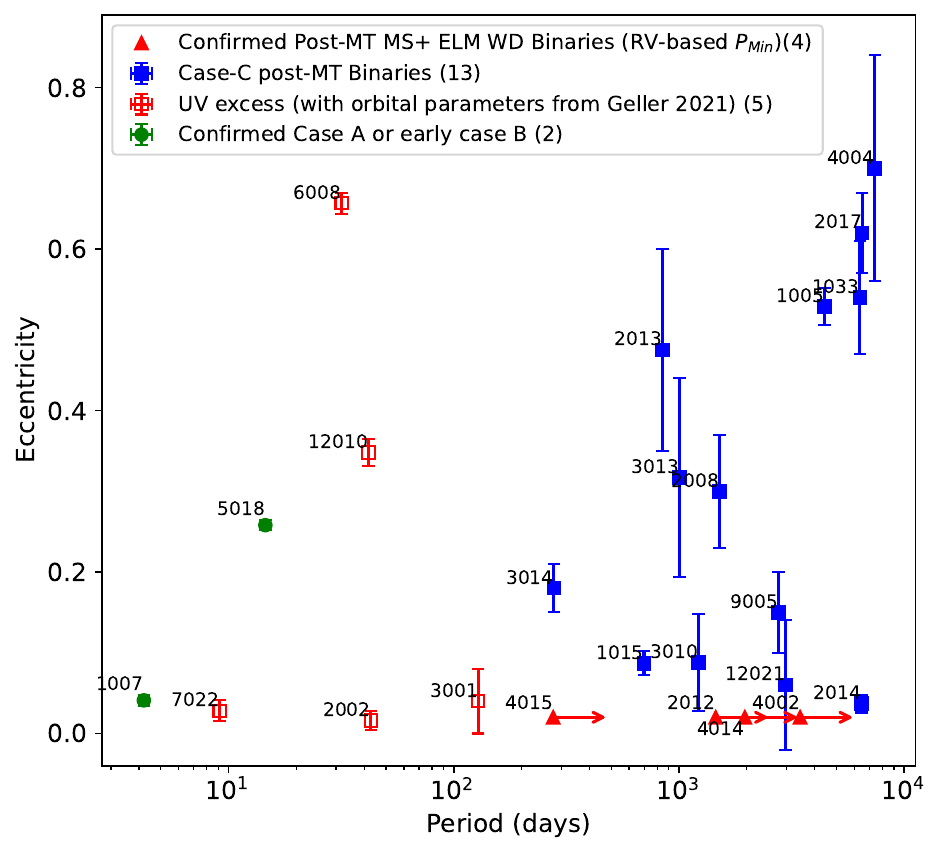}
    \caption{Period vs eccentricity (with error bars)  of binary systems with MT scenario classification with their WOCS IDs}
    \label{fig:Period}
\end{figure}

Fig~\ref{fig:Period} presents the period vs. eccentricity plot of binary systems, with Case~A or Case~B and Case~C MT, as mentioned in Table~\ref{tab:params} and Table~\ref{tab:abundance}. The plot clearly illustrates the transition between these MT scenarios, highlighting how MT in binaries influences the orbital period of these systems.

\end{appendix}

\end{document}